\documentclass[acmsmall]{acmart}
\AtBeginDocument{%
  }

\setcopyright{acmlicensed}
\copyrightyear{2026}
\acmYear{2026}
\acmDOI{XXXXXXX.XXXXXXX}

\acmJournal{JACM}
\acmVolume{37}
\acmNumber{4}
\acmArticle{111}
\acmMonth{8}

\begin{document}

\title{Beyond the Silence: How Men Navigate Infertility Through Digital Communities and Data Sharing}

\author{Tawfiq Ammari}
\affiliation{%
  \institution{Rutgers University}
  \city{New Brunswick, NJ}
  \country{USA}}
\email{tawfiq.ammari@rutgers.edu}

\author{Zarah Khondoker}
\affiliation{%
  \institution{Johns Hopkins Bloomberg School of Public Health}
  \city{Baltimore, Maryland}
  \country{USA}}
\email{zkhondo1@jh.edu}

\author{Yihan Wang}
\affiliation{%
  \institution{University at Buffalo}
  \city{Buffalok, NY}
  \country{USA}}
\email{ywang492@buffalo.edu}

\author{Nikki Roda}
\affiliation{%
  \institution{Rutgers University}
  \city{New Brunswick, NJ}
  \country{US}}
\email{nikki.roda@gmail.com}
\renewcommand{\shortauthors}{Ammari et al.}

\begin{abstract}
Men experiencing infertility face unique challenges navigating Traditional Masculinity Ideologies that discourage emotional expression and help-seeking. This study examines how Reddit's r/maleinfertility community helps overcome these barriers through digital support networks. Using topic modeling (115 topics), network analysis (11 micro-communities), and time-lagged regression on 11,095 posts and 79,503 comments from 8,644 users, we found the community functions as a hybrid space: informal diagnostic hub, therapeutic commons, and governed institution. Medical advice dominates discourse (63.3\%), while emotional support (7.4\%) and moderation (29.2\%) create essential infrastructure. Sustained engagement correlates with actionable guidance and affiliation language, not emotional processing. Network analysis revealed structurally cohesive but topically diverse clusters without echo chamber characteristics. Cross-posters (20\% of users) who bridge r/maleinfertility and the gender-mixed r/infertility community serve as navigators and mentors, transferring knowledge between spaces. These findings inform trauma-informed design for stigmatized health communities, highlighting role-aware systems and navigation support.
\end{abstract}



\received{20 September 2025}
\received[revised]{12 December 2025}
\received[accepted]{15 January 2026}

\maketitle

\section{Introduction}
Male infertility accounts for 20–30\% of infertility cases \cite{agarwal2015unique}, yet remains under-discussed \cite{patel_et_al_2019,arya2016experience}, under-supported \cite{costa_et_al_2017,patel_et_al_2019}, and heavily stigmatized \cite{gannon_masculinity_2004,dooley2011psychological}. Beyond thwarting reproductive goals, it challenges cultural linkages between masculinity and virility, introducing biological, relational, and psychological uncertainties while offering few social scripts or institutional resources for coping. Unlike many areas of reproductive health that benefit from gendered solidarity and advocacy, men experiencing infertility often face profound isolation. In the absence of formal or culturally accessible support systems, many turn to online spaces for guidance and connection. 

Reddit, a predominantly male platform \cite{Pew2024SocialMedia}, has become a critical site for masculine identity (re)construction \cite{laviolette_using_2019,farrell_et_al_2019} as socio-economic changes threaten men's traditionally privileged status \cite{hanson2023s}. This is especially evident in male infertility discussions, where men feel "a part of their masculinity is being taken away" \cite{patel_et_al_2019}, and stigma is so pervasive that men have opted to conduct research interviews in parked cars to consciously ensure their privacy \cite{dolan_its_2017}. Reddit's pseudonymity enables engagement with stigmatized topics \cite{ammari_et_al_19,de2014mental}, fostering solidarity networks \cite{morini2025participant} that protect against depression and suicidal ideation while challenging hegemonic norms of stoicism \cite{redondo2021create}. The platform's topic-specific communities (subreddits), moderated by peers facing similar challenges \cite{Gilbert_23,naslund2016future,proferes2021studying} and governed by community-specific rules \cite{fiesler2018reddit}, enable active support spaces \cite{morini2025participant}. This study focuses on \href{https://www.reddit.com/r/maleinfertility/}{r/maleinfertility}, described as a space \textit{"for men experiencing infertility and male perspectives on infertility. Partners are encouraged to participate, but we ask they post in the daily Partner's Perspectives thread."}

HCI and CSCW research shows how online health communities foster emotional expression, identity work, and peer support \cite{ammari_et_al_2018,andalibi_forte_18,andalibi_2020}, providing "social transition machinery" \cite{haimson_18} that helps people understand \cite{Ammari2024} and heal from trauma \cite{Randazzo2025}. Studies of stay-at-home fathers reveal that men use online forums to reframe vulnerability as compatible with strength \cite{ammari_et_al_2017,ammari_et_al_19,lupton2016parenting}, sometimes creating father-only spaces \cite{ammari_stayhome_dads_16}. Yet—except for Patel et al. \cite{patel_et_al_2019}—male infertility has received little attention in social computing scholarship. We therefore ask:
\begin{quote}
\textbf{RQ1a:} What kinds of discourse emerge in male infertility communities on Reddit?
\end{quote}

While female infertility discourse is well theorized \cite{patel_et_al_2019}, male infertility remains underexplored. Understanding its emotional and linguistic structures can inform platform design, given that user retention is central to online support community health \cite{wan_et_al_20,jo_et_al_23}. Prior work suggests affective strategies—sharing vulnerability, mentoring, identity reconstruction—boost engagement \cite{Ammari2024,andalibi_2020,vermeeren2025haunting}, but it is unclear which forms of self-expression are most rewarded in male infertility communities. Addressing this lack of clarity regarding male-centric support, we ask:
\begin{quote}
\textbf{RQ1b}: What types of discourse and user attributes lead to higher engagement in male infertility communities?
\end{quote}

Reddit operates across macro, meso, and micro normative levels \cite{Chandrasekharan_et_al_2018,cai_wohn_21,ammari_et_al_22,vachler_et_al_2020}, shaping moderation and acceptable discourse in ways that vary even within subreddits \cite{weld2024making}. Users select communities based on rules \cite{zou2024self,proferes2021studying}, but engage with posts according to personal interest \cite{yu_et_al_24,zhang2017community}. In stigmatized health contexts, while homophilous groups offer vital support \cite{randazzo_ammari_2023,randazzo_et_al_23}, users may encounter narrow perspectives dominated by despair narratives \cite{Randazzo2025,zhong2024impact,joy_et_al_25}. These dynamics can produce \textit{grief bubbles}—echo chambers where users risk retraumatization through repeated exposure to similar discussions without encountering alternative perspectives \cite{Randazzo2025,goglia2024structure,mancini2022self}. Drawing on definitions of echo chambers as combining structural closure with topical purity \cite{morini2021toward}, we ask:
\begin{quote}
\textbf{RQ2a}: Are there distinct sub-communities (echo chambers) within the male infertility subreddit?
\end{quote}

Online community members adopt distinct functional identities—experts, translators, facilitators \cite{saxena2021users,Kou_et_al_18}. Access to multiple roles proves crucial for identity reconstruction \cite{laumann_bonds_1973}, as individuals navigating life transitions require varied relational support. Ammari et al. \cite{Ammari2024} demonstrate this multiplicity in foster care communities, where members simultaneously identify as parents, survivors, and professionals—role multiplicity enabling identity reconstruction across temporal stages and relational positions. To examine how this role multiplicity manifests in our context, we ask:
\begin{quote}
\textbf{RQ2b}: What roles are represented within the different sub-communities in the male infertility subreddit?
\end{quote}

The appeal of joining communities stems from discovering one's "people" \cite{TeBlunthuis_el_al_22}. Homophilous clustering creates protective environments for processing trauma across domains like mental health \cite{pavalanathan_identity_2015}, sexual assault \cite{andalibi_haimson_decchaudhury_forte_2016}, parenting \cite{ammari_et_al_19,ammari_et_al_2017,ammari_stayhome_dads_16}, caregiving \cite{ammari_et_al_2015_networked}, child loss \cite{andalibi_giniel_2022}, and foster care \cite{Ammari2024}. Participation across multiple communities enables tailored self-presentation \cite{litt_imagined_2016,TeBlunthuis_el_al_22}. While cross-posters can be hostile elsewhere \cite{kumar2018community}, in trauma-affected domains they serve as knowledge brokers and empathy-builders \cite{Ammari2024}. In male infertility contexts, cross-posting between male-centric and broader communities may expand resources and normalize emotional expression. To understand the nature of interactions at these community intersections, we ask:
\begin{quote}
\textbf{RQ3}: What are the characteristics and functions of users who cross-post between male infertility subreddits and broader infertility communities?
\end{quote}

By addressing these questions, we extend Patel et al. \cite{patel_et_al_2019} through large-scale analysis of male infertility communities and contribute to HCI conversations on trauma-informed design, digital support infrastructures, and gendered online identity work. Our findings aim to inform the design of more inclusive, empathetic platforms for reproductive health support, particularly for conditions that remain invisible or culturally silenced.

\section{Related Work}
In this section, we locate our study at the intersection of gender, online support, and design. We first review how Traditional Masculinity Ideologies (TMI) shape men’s coping and help-seeking, highlighting how infertility constrains identity and disclosure. We then examine how online communities foster connection and identity (re)construction through personal data sharing and peer interaction. Finally, we draw on trauma-informed design to consider platform affordances—pseudonymity, algorithmic curation, and moderation—that enable or hinder safety, agency, and peer support. Together, these strands explain why men turn to Reddit, how support practices form, and which design choices sustain engagement in male-infertility communities.

\subsection{The Role of Traditional Masculinity Ideologies in Men's Mental Health}
\label{sec:related_work_TMI}
Men often show distinct behavioral patterns, including higher rates of aggression \cite{Bjorkqvist2018}, substance misuse \cite{McHugh2018}, and suicide \cite{Coleman2020}. These patterns are partly rooted in Traditional Masculinity Ideologies (TMI), which valorize self-reliance, emotional restraint, and dominance \cite{Messerschmidt2019}. By discouraging emotional expression, TMI fosters harmful coping strategies like alcohol use and unmanaged stress \cite{Cleary2012,Oliffe2023,Tsan2011}, which can escalate into aggression, social withdrawal, and suicidal behavior \cite{Krysinska2010}.

Men's emotional expression is context-dependent. Limited social settings like funerals or sporting events can permit displays of emotion without threatening masculine norms \cite{Walton2004}, yet pervasive cultural messages such as “boys don’t cry” continue to suppress emotional development \cite{Frosh2003,Creighton2013}. Dolan et al. \cite{dolan_its_2017} found that men facing infertility often avoided sharing their struggles with friends or partners to escape ridicule, adopting a “perverse passivity” in which infertility was something endured rather than addressed. Our study explores whether digital spaces—offering anonymity and asynchronous interaction—can create alternative contexts that legitimize emotional expression.

Gender differences are also evident in help-seeking. Women are significantly more likely than men to engage with online mental health therapy \cite{Titov2017}, and men who begin therapy frequently drop out due to stigma, vulnerability, or concerns about autonomy—barriers reinforced by TMI \cite{EnglarCarlson2006,Aguera2017}. These obstacles compound psychological distress and encourage maladaptive coping \cite{Seidler2016}.

Within infertility care, help-seeking is further constrained by the framing of infertility as a female-centered issue, with male-centered discussions focusing narrowly on sperm quality and count \cite{gannon_masculinity_2004}. Gannon observed that declining sperm counts were depicted as a “war” against men’s bodies, reinforcing narratives of emasculation or quasi-castration. As Patel et al. note \cite[P.10]{patel_et_al_2019}, such portrayals can deter men from seeking help—whether consulting doctors, confiding in friends, or even talking with partners. If virility and reproduction are central TMI ideals, infertility relegates men, even those otherwise “normal,” to the margins of masculinity.

Traditional masculine ideology—emphasizing self-reliance, emotional restraint, and stoicism—ostracizes men who disclose trauma or support vulnerable community members, creating barriers to help-seeking \cite{mokhwelepa2025men}. Semaan et al.'s studies of U.S. military veterans demonstrate how "hyper-masculinity" learned during service impedes post-service disclosure and civilian reintegration \cite{semaan_et_al_16,semaan_et_al_17}, leading veterans to engage in "hyper-selective disclosure" through information communication technologies \cite{dosono_et_al_17}. These norms shape disclosure even in digital environments, suggesting men with stigmatized health conditions may similarly restrict online help-seeking. Complicating matters further, men perceiving themselves as less masculine report higher harassment endorsement, mediated by masculine norm adherence and toxic disinhibition \cite{rubin_et_al_20}. However, this does not preclude finding support online—communities can support marginalized men, such as transmen navigating identity-related crises \cite{dym_et_al_19}, facilitating identity work difficult in offline contexts constrained by normative expectations.

\subsection{Digital Pathways to Social Connection, Belonging, and Well-being in Men’s Mental Health}
\label{sec:related_work_information_support}
Social connectedness is critical to physical and mental health; its absence raises risks of cardiovascular disease and early mortality \cite{Bailey2018,HoltLunstad2022}. Men may be especially vulnerable because their social networks are often smaller and less emotionally expressive \cite{Steptoe2013}. In this context, digital platforms have emerged as key venues for connection and support in diverse areas including fatherhood \cite{ammari_et_al_19,ammari_et_al_2017,sepahpourfard_mommit_daddit_22}, pregnancy loss \cite{andalibi_2019,andalibi2018social}, parenting children with special needs \cite{ammari_et_al_2015_networked}, foster care \cite{lee2025artificial,Ammari2024,LEE2021105262,fowler_former_foster_youth_22}, and mental health \cite{morini2025participant,de_choudhury_predicting_2013,de2014mental,saha_et_al_20}. We focus on three central functions of these communities: (1) building collective knowledge and informal expertise; (2) care work; and (3) (re)building identity and belonging. 

\subsubsection{Personal Data Sharing as Collective Knowledge and Informal Expertise}
\label{sec:related_work_data_sharing}
Online support communities connect people with shared experiences \cite{fox_social_2014,rainsSocialSupportComputerMediated,MechanismsLinkingSocialTiesandSupporttoPhysicalandMentalHealth2011}, offering critical insights to peers and caregivers \cite{schorch_designing_2016,SocialNetworksExperientially2013}. While individuals weigh benefits and drawbacks of data sharing \cite{wu_effects_2023}, intrinsic rewards like self-worth and satisfaction motivate ongoing participation \cite{yanKnowledgeSharingOnline2016,SocialCapitalIndividual2011}.

Self-tracking is common in managing chronic conditions \cite{10.1145/1753326.1753409}, yet sharing such data can be emotionally fraught \cite{knittel_anyone_2021}. Through community engagement, posting personal metrics provides emotional support, celebrates progress, solicits information, and fosters accountability \cite{chungWhenPersonalTracking2017,guiWhenFitnessMeets2017,luModelSociallySustained2021a}, helping users reinterpret data, validate experiences, and reinforce coping mechanisms \cite{knittel_anyone_2021}. 

Patel et al. similarly found men sharing detailed fertility regimens and digital narratives of their infertility journeys \cite[P.9]{patel_et_al_2019}.
\textit{Informational support}—knowledge, advice, and guidance—helps people navigate health challenges through peer support offering experiential knowledge \cite{gage2013social,macgeorge2004myth} that complements professional care \cite{levonian_et_al_21}. This enables "collective sensemaking": collaboratively interpreting health information to co-construct shared understanding \cite{mamykina_collective_2015,OnlinePeerSupport2023} while normalizing experiences, reducing uncertainty, and alleviating shame \cite{choe_understanding_2017}. 

\subsubsection{Care Work in Online Health Communities}
\label{sec:related_work_carework_ohc}
MacGeorge et al. \cite{macgeorge2004myth} identified three additional types of support in online health communities. \textit{Emotional support} offers warmth, empathy, and encouragement through spaces for venting and validation—Progga et al. \cite{progga_understanding_2023} found women with postpartum depression seeking assurance that their emotions were normal, while women experiencing infertility similarly gained support through venting in safe spaces \cite{zou2024self}. \textit{Esteem support} communicates respect and validation; Payne et al. \cite{payne_building_2025} documented fat individuals building solidarity through mutual validation to resist medical stigma, while older members of feminist communities gained standing as veterans with valuable experience \cite{ammari_et_al_22}. \textit{Network support} cultivates belonging through community ties, reducing isolation and enabling collective action \cite{ammari_et_al_2015_networked}—members support newcomers by inviting direct messages and establishing new social connections \cite{de2014mental,ammari_et_al_19,Ammari2024}.


\subsubsection{(re)Building Identity and Belonging to a Community}
\label{sec:related_work_rebuilding_ID}
\textit{Network support} is vital because belonging protects against depression and suicidal ideation \cite{McLaren2009}, yet strategies for cultivating it among isolated men are limited. Digital forums offer connection through shared experience \cite{Biagianti2018}, and well-designed digital interventions can match face-to-face treatment efficacy \cite{Hull2020,Rahman2022,Leite2019}.

Online health communities enable patients and caregivers to provide mutual support \cite{levonian_et_al_21,joy_et_al_25}, helping men—who often lack offline networks—combat isolation \cite{rains_coping_2024} while serving as essential organizing spaces for marginalized groups facing stigmatized conditions \cite{payne_building_2025,progga_understanding_2023}. Men explicitly request support in these settings \cite[pp.8-10]{patel_et_al_2019}. Yet gender shapes engagement: Zou et al.'s \cite{zou2024self} analysis of Reddit's IVF community found women maintaining "close yet detached" relationships through "tenuous ties," with informational support predominating while esteem, network, and tangible support remained rare. Women used the subreddit primarily for venting (emotional support) and \textit{information support} rather than lasting connections, suggesting online communities may function differently across gendered infertility experiences. Such variation reflects platform affordances and moderation norms. Designing effective spaces requires trauma-informed principles \cite{randazzo_et_al_23} that respect user boundaries, avoid forced disclosure or stigma reproduction, and support agency through safe expressive modes \cite{Ankrah_et_al_2022}. We discuss these considerations next.

\subsection{Trauma-informed Design}
\label{sec:related_work_trauma_informed}
Trauma-informed design frameworks \cite{chen_trauma-informed_2022,Scott2023} highlight how social media affordances—such as algorithmic recommendation and comment sorting \cite{parchoma_contested_2014}—and governance practices like moderation \cite{Jhaveretal2018} can support vulnerable populations. Three core principles guide this work: peer support, user choice, and safety, all underpinned by transparency and trust \cite{Bellini2023,dym_fiesler_2020}. Adhering to these principles fosters identity reconstruction and psychological resilience through safe environments for trauma recovery and emotional validation \cite{Ammari2024,Randazzo2025}.

\subsubsection{Peer support: Building supportive online spaces} \label{sec:peer_support_trauma_informed}
Trauma-informed design centers shared understanding and mutual aid \cite{chen_trauma-informed_2022,Scott2023}. Peer support is shaped by feedback mechanisms: direct signals like upvotes and replies communicate community norms \cite{laviolette_using_2019}, while observing others’ posts also teaches these norms \cite{randazzo_ammari_2023}. Such signals—positive or negative—serve as barometers of community endorsement.

Algorithmic recommendation can enhance transportability, exposing users to narratives that resonate with their experiences which supports peer identification \cite{randazzo_ammari_2023,green2021transportation}. Yet algorithmic opacity poses risks: emotional content may be amplified to drive engagement, harming well-being and limiting community control \cite{matias2019preventing,Jhaveretal2018,Fabbri_23}.

\subsubsection{Pseudonymity, Algorithmic Curation, and Choice} \label{sec:user_choice_voice}
Reddit’s pseudonymity and limited identity cues facilitate disclosure of stigmatized experiences such as sexual assault, trauma, family struggles, and eating disorders \cite{andalibi_2019,andalibi2018social,haimson_disclosure_2015,randazzo_ammari_2023,chancellor_norms_2018}. Users exercise agency by selecting communities of interest \cite{Huang_Foote_21}, cross-posting content \cite{datta2017identifying}, and managing overlapping networks of support \cite{Ammari2024}. Platform features like badges, awards, and peer recognition further influence participation: awards encourage newcomers, while badges help sustain engagement among committed members \cite{burtch2022peer,trujillo2022assessing,bornfeld2017gamifying} -- both of which are important given that retention drives online support community health \cite{wan_et_al_20,jo_et_al_23}

\subsubsection{Moderation and community governance}
\label{sec:related_work_mod}
Online community safety depends on moderation that defines acceptable discourse,  harmful content, and bans disruptive users \cite{Jhaveretal2018,mansour_et_al_21}. While automated tools assist at scale \cite{gillespie2018custodians,jhaver2019human}, they miss contextual nuance in emotionally complex content because harm definitions are context-specific \cite[p.98]{gillespie2018custodians,ammari_et_al_22}, and algorithmic opacity erodes trust when decisions remain unexplained \cite{kuo_et_al_23}. Human moderators therefore perform essential labor: shaping discussion boundaries, organizing information, applying community norms \cite{edwards_moderator_2002}, and defining membership—as in fathers-only groups excluding mothers \cite{ammari_stayhome_dads_16,ammari_et_al_19}. This constitutes complex care work across multiple power levels, particularly when providing supportive spaces for disclosure while protecting privacy around stigmatizing issues \cite{saha_et_al_20,ammari_et_al_22}. Drawing on Collins' matrix of oppression, Gilbert \cite{Gilbert_23} shows how r/AskHistorians moderators engage in intensive emotional labor at personal, community, and systemic levels while contending with platform policies and sociotechnical constraints enabling harmful content. This boundary work extends to active advocacy: Wu et al. \cite{wu_et_al_24} demonstrate how city subreddit moderators establish institutional, cultural, and geographical boundaries to counter racist discourse and brigading. Crucially, moderation functions as both protection and collective empowerment—Ammari et al. \cite{ammari_et_al_22} show moderators of women-only spaces acting simultaneously as gatekeepers and advocates empowering members to challenge societal norms—revealing moderation as relational and emotional labor that defends boundaries while fostering community agency.

Our study extends this framework to stigmatized populations—men experiencing infertility—examining how moderation tactics protect members from content or actors that could compound medical and emotional distress or violate the trust necessary for vulnerable self-disclosure.

\section{Methods}
This study is part of a broader project on men’s traumatic experiences online. We began with a two-month survey of eleven infertility-, parenting-, adoption-, and related subreddits to map discursive patterns and identify communities of interest. All four team members read posts, took notes, and exchanged memos.

Focusing on male infertility, we collected all posts and comments from r/maleinfertility via the PushShift API \cite{baumgartner2020pushshift} and the Arctic Shift dataset \cite{heitmann_arctic_shift_2025}, yielding 79,503 comments and 11,095 posts by 8,644 users. Because roughly 20\% of these users (1,725) also post to r/infertility, we analyzed both community dynamics in r/maleinfertility and characteristics of these cross-posting users.

Our approach follows Tornberg and Tornberg’s mixed-methods framework \cite{tornberg2016combining}, combining computational modeling with qualitative interpretation. Qualitative analysis guided topic selection, model refinement, and interpretation.

\subsection{Topic Model Training and Parameter Tuning (RQ1a)} \label{topic_mod}
To answer RQ1a, We applied BERTopic \cite{grootendorst_bertopic_2022} to Reddit threads treated as documents, using BERT embeddings to cluster semantically similar content \cite{reimers_sentence-bert_2019,angelov_top2vec_2020}. To optimize performance we trained 32 models, selecting the best via u\_mass coherence \cite{newman2010automatic,roder_exploring_2015,chang_reading_2009}. The top configuration (coherence = –0.79) produced 115 topics. For each comment we assigned the dominant topic based on BERTopic’s probability distributions \cite{mcinnes_hdbscan_2017,campello_density-based_2013}. Full hyperparameter details appear in Appendix \ref{appendix:topic_model}.

\subsubsection{Qualitative coding}  \label{sec:method_bertopic_qualitative}
We sampled 15 comments per top topic (including parent threads). Each topic was double-coded by two authors, with regular meetings to define topic boundaries and select exemplar quotes, following consensus methods \cite{mcdonald_et_al_19}. Through axial coding \cite{wall1999sentence} we identified 14 cross-topic themes—e.g., \textit{Advice on Insurance, Doctors, and Medication Expense; Comfort Posting; Partner Emotional Support}—which we grouped into three meta-themes: Medical Advice, Emotional Support, and Moderation (see Appendix \ref{appedix:topic_breakdown}).

\subsection{Time-lagged Regression: Future Use Predictors (RQ1b)} \label{sec:time_lag_reg_meth}
To answer RQ1b, we analyzed comment-level records from r/maleinfertility, containing a user identifier and a timestamp.  Events were binned into monthly periods. We operationalized community activity as the number of comments per user per month. To account for temporal dynamics, we created lagged predictors that reflected each feature’s value in the prior month. The dependent variable was defined as the number of comments contributed in the subsequent month ($t+1$). Only users with at least two months of data were included.

In earlier CHI and CSCW work, time-lagged regression was used to uncover temporal dynamics in digital behavior. Mitra et al.\cite{mitra_et_al_17} applied it to model how employee engagement spreads across a corporate social network, using prior engagement levels of an individual’s contacts as lagged predictors of their future engagement. Li et al. \cite{Li_et_al_22} used time-lagged regression to predict the emergence of social media gatekeepers on Twitter by linking users’ past behaviors and profile features to their future influence in immigration-related news dissemination. In both cases, time-lagged models enabled the researchers to infer directional and predictive relationships over time.

\subsubsection{Feature Set Permutations} \label{sec:features}
We compared models using user metadata, LIWC categories, and thematic weights (\S\ref{sec:method_bertopic_qualitative}). The best specification (RMSE = 2.88, $R^2$ = 0.66) combined all three.

\paragraph{User metadata (11 variables)}: average comment length and sentiment (VADER \cite{elbagir2019twitter,hutto2014vader}); number of responses and karma per comment \cite{laviolette_using_2019}; response probability; whether a comment was edited; average response length, sentiment, and karma; time to first response; and total responses received.

\paragraph{LIWC (70 features)}: lexical and syntactic categories validated for social media \cite{pennebaker_linguistic_2001,pennebaker_linguistic_2007,de_choudhury_predicting_2013}.

\paragraph{Themes (14 features)}: mean topic weights aggregated into the qualitative themes (see \S\ref{sec:themes}).

We standardized all predictors and used L1 (Lasso), L2 (Ridge), and Elastic Net regularization to reduce dimensionality and multicollinearity.


\subsection{Studying interactions within the r/maleinfertility community:Network Analysis (RQ2)}
\label{sec:meth_network}
To answer RQ2, we built a directed, weighted reply network where nodes are users and edges represent replies. Edges were aggregated by interaction count, self-loops removed, resulting in 8,536 nodes and 53,081 edges.

\subsubsection{Community detection}
We symmetrized edge weights and applied the Louvain algorithm to maximize modularity \cite{newman2018networks,cohen2020power}.
We refer to each of the micro-communities in this dataset as clusters in our analysis to reduce any confusion with the r/maleinfertility community. 

\subsubsection{Cluster Structural and Linguistic Analysis}
\paragraph{Conductance (structural insulation)}
The conductance $C$ of a cluster is defined by the total volume of edges leaving it as a ratio of the total edges:
\[
\phi(C) \;=\; \frac{W_{\text{out}}(C)}{\,W_{\text{in}}(C) + W_{\text{out}}(C)\,}\;\in[0,1].
\]
Lower $\phi$ implies stronger internal cohesion.

\paragraph{Reciprocity in Networks}
captures the proportion of mutual ties within a cluster \cite{morini2021toward}

\[
r = \frac{\sum_{i \neq j} A_{ij} A_{ji}}{\sum_{i \neq j} A_{ij}}.
\]

\paragraph{Purity}
is the fraction of nodes sharing the most common stance label (theme × sentiment) \cite{morini2021toward,ALDAYEL2021102597,naskar2016sentiment}

\[
P_c = \prod_{a \in A} \max \left( \sum_{v \in c_a} (v) \right)
\]

\paragraph{Theme entropy}
quantified within-cluster topical diversity using Shannon entropy:
\[
H(C) \;=\; -\sum_{t} p(t\mid C)\,\log p(t\mid C).
\]
Lower $H(C)$ implies a narrow topical focus (content homogeneity), while higher $H(C)$ indicates more diverse subject matter. 

\subsubsection{Detecting Echo Chambers} \label{subsec:echo_meth}
Following Morini et al. \cite{morini2021toward}, we define echo chambers as clusters with Conductance < 0.5 and Purity > 0.7. To define purity, we start by labeling user stance which we defined as (Qualitative Theme | VADER sentiment). In other words, for each theme, a stance would be (theme X positive; theme X negative; and theme X neutral). This is in keeping with earlier work on stance as a product of topic and sentiment \cite{ALDAYEL2021102597,naskar2016sentiment}. We used this method to answer RQ2a.

\subsubsection{Qualitative Analysis of Clusters}
\label{sec:meth_qual_cluster}
For each Louvain cluster, we sampled 10 comments from the most central users (by PageRank), double-coded them, and met to reach consensus on cluster character and central user roles. Initial categories (\textit{mentor, moderator, expert}) were refined to reflect nuanced behaviors exhibited by those roles such as \textit{system critic} and \textit{newcomer}. We used this method to answer RQ2b.

\subsection{Classifier: Who are the cross-posters? (RQ3)}
\label{sec:meth_cross_posting}
We framed cross-posting as a binary classification task with target \textit{cross-post} $\in$ {0,1} representing whether the user posted to both r/maleinfertility and r/infertility. 

We used the same feature set described in \S\ref{sec:features}.
We framed cross-posting (posting to both r/maleinfertility and r/infertility) as a binary classification task. Using the same 95 features, we trained and evaluated Logistic Regression, Random Forest, Extra Trees, Gradient Boosting, HistGradientBoosting, AdaBoost, GaussianNB, KNN, LinearSVC, and XGBoost/LightGBM with 5-fold StratifiedKFold. The best model achieved accuracy = 0.846, precision = 0.793, recall = 0.307, F1 = 0.442, and ROC–AUC = 0.835, indicating strong fit \cite{lemeshow_review_1982,fawcett_introduction_2006,swets_signal_2014}. We applied SHAP to the final Gradient Boosting model to identify the most predictive features \cite{shaplundberg2017unified}. Each SHAP value indicates a feature’s additive effect on the log-odds of cross-posting, with positive values increasing and negative values decreasing the likelihood of cross-posting \cite{shaplundberg2017unified}. This method allowed us to answer RQ3.


\subsection{Positionality and Ethical Stance} \label{subsec:ethical_stance}
Our research team is committed to supporting individuals navigating trauma, uncertainty, and life transitions. The first author, a cisgender man, occupies partial insider status—sharing gender identity with participants while maintaining analytical distance—and studies how social technologies simultaneously empower vulnerable communities and pose risks to privacy, autonomy, and identity. The second author, a cisgender South Asian woman and graduate student in reproductive and public health, grounds the study in clinical and epidemiological context essential for interpreting community discussions of diagnostics and treatment. The third and fourth authors—cisgender East Asian and white Latinx women respectively—bring critical external perspectives on masculinity construction; the fourth author's specialization in surfacing human stories behind complex datasets bridges our computational and qualitative methods.

Following AoIR Ethical Guidelines \cite{franzke2020internet}, we mitigated risks of weaponization by bad actors. Because direct quotes might contain sensitive information usable against individuals \citep{sloan2022sage}, and online users often do not expect to be quoted in research \cite{fiesler_participant_2018,Fiesler_et_al_24}, we paraphrased community stances and edited supporting quotes to protect anonymity \cite{bruckman2002studying,Markham01042012}, preventing digital traceability while preserving original intent. \textit{Disguised quotes appear italicized throughout the paper for clarity.}

\section{Findings}
Our findings unfold across three layers of analyses. In \S\ref{sec:themes}, we answer RQ1a by presenting a thematic analysis showing how medical advice, emotional support, and moderation position the subreddit as a diagnostic hub, therapeutic commons, and digital institution. We go on to answer RQ1b in \S\ref{sec:future_engagement} by describing what influenced future community engagement. In \S\ref{sec:cluster_findings}, to answer RQ2, we examine community structure as well as user roles—mentors, experts, hope brokers, and system critics—along with how moderation and partner participation shape collective identity. Finally, to answer RQ3, \S\ref{sec:cross_posting} explores cross-posting, revealing how users who bridge r/maleinfertility and r/infertility serve as navigators and mentors, aligning norms across spaces and reinforcing male-centered support.

\subsection{Diagnostic Utility Over Emotional Support: Themes Driving Future Engagement (RQ1a)} \label{sec:themes}
Our analysis of 115 discussion topics, grouped under three overarching themes—\textbf{Medical Advice, Emotional Support, and Moderation}—illustrates how the infertility community on Reddit blends biomedical knowledge, affective support, and digital governance. Medical Advice (63.3\%) centers on interpreting diagnostics, navigating treatment, confronting biological challenges, and addressing systemic barriers like insurance. Emotional Support (7.4\%) encompasses grief, coping rituals, comfort posting, and strategies for resilience. Moderation (29.2\%) highlights onboarding, navigation aids, rule enforcement, and wiki/FAQ infrastructure. Together, these themes reveal how the community functions simultaneously as an informal diagnostic hub, therapeutic commons, and regulated digital institution—providing an antidote to the isolation many men experience during infertility. A detailed breakdown appears in Appendix \S\ref{appedix:topic_breakdown}.

\subsubsection{Medical Advice: Navigating Diagnostics, Treatments, and Systemic Barriers} \label{subsec:mecical_advice}
The most prominent theme centers on medical advice and information exchange, where users interpret results, evaluate treatments, and contextualize biological and systemic challenges. Given male infertility's diverse causes, discussions focus on problem-solving: how do I overcome this specific problem?

\paragraph{Medical Report Analysis}
Facing research gaps that systematically exclude men from fertility studies \cite{harlow2020qualitative}, users described \textit{"grasping at straws"} for male-factor information. Peers filled this void by collaboratively decoding semen analyses \cite{gannon_masculinity_2004}, reassuring those who feared flagged numbers meant they were \textit{"doomed"} by reframing metrics contextually—explaining that percentages mislead when counts are high and that 0\% morphology (shape and size of semen)\footnote{Sperm morphology refers to the size and shape of sperm, which is one factor analyzed in a semen test to evaluate male fertility -- Source: \url{https://www.mayoclinic.org/diseases-conditions/male-infertility/expert-answers/sperm-morphology/faq-20057760}} does not preclude conception.

\paragraph{Health and Well-being}
Beyond treatment, members share lifestyle strategies: supplements like vitamin E as well as advice to \textit{``ditch the laptop on the lap''} or \textit{``switch to cooler underwear.''} Though anecdotal, these exchanges position participants as active agents. Some share specific routines—workout timing, masturbation frequency, diets—crowdsourcing and normalizing conception efforts.

\paragraph{Advice on Insurance, Doctors, and Medication Expense}
Financial navigation forms another critical sub-theme. Posters swap tips on discount programs (e.g., Compassionate Care), employment benefits (e.g., Starbucks part-time IVF coverage), and shared-risk refund schemes. Others compare international systems: \textit{``in the Netherlands, three cycles are covered with a deductible''} versus U.S. cycles often exceeding \$10,000 out-of-pocket. This highlights inequities where access depends on geography and insurance literacy as much as biology. Users advocated for at-home testing kits when clinic access was limited, but this self-advocacy was complicated by labs repeatedly rejecting samples. One user lamented the discouraging cycle of awkwardly producing multiple specimens only to face continued rejection.

\paragraph{(Fe)male Infertility Generalized Guidance: Navigating Infertility and Partner Dynamics}
These sub-themes reveal Reddit as a secondary diagnostic and logistical space where members crowd-source interpretations and trade experiential knowledge, filling gaps left by inaccessible medical care. Threads span female concerns—ovulation, endometrial preparation, embryo transfer—and male diagnoses like azoospermia\footnote{Condition where a man's ejaculate contains no sperm — Source \url{https://my.clevelandclinic.org/health/diseases/15441-azoospermia}} and low motility. One user described treatment as \textit{``another full-time job''}; others expressed frustration with medications improving hormone levels while sperm counts remained zero—\textit{``heartbreaking to fix the chemistry on a chart but still see no change.''} Yet these informational exchanges carry emotional weight. Partners describe male factor infertility transforming \textit{``my problem into our battle,''} with one writing: \textit{``I would rather have this husband and no baby than a baby and no husband.''} Men echoed this, framing relationships as a \textit{``trust fall''}—partners must support each other's grief without hiding their own pain. Some tensions emerged, with moderators asking partners to avoid being \textit{``all doom and gloom''} and stick to medical facts, recognizing men already face significant pressure. As these examples suggest, medical discussions often blend into emotional territory—a theme we explore more fully in the following section.



\subsubsection{Emotional Support: Coping, Resilience, and Connection} \label{subsec:findings_emotional_support}
Running parallel to biomedical exchange is a robust theme of \textit{emotional support}, where the subreddit functions as a therapeutic scaffold. Members here actively provide emotional care work (see \S\ref{sec:related_work_carework_ohc}) such as acknowledging that dealing with \textit{"miscarriages and infertility is miserable - all of your feelings are valid.}"  Words of encouragement to \textit{"accept yourself in your disease"} but not be \textit{"defined"} by it are often repeated. Newcomers are encouraged to ask many questions as they want, and to DM and reach out to veteran members of the community. 

\paragraph{Grief and Disappointment} 
Threads feature raw narratives following miscarriage, failed transfers, or prolonged infertility. Posters describe feeling \textit{"numb,"} \textit{"burned out,"} or \textit{"going through the motions while emotionally checked out."} Community replies emphasize validation—reminding members that every loss, even a chemical pregnancy,\footnote{Pregnancies ending early, around the fifth week -- Source: \url{https://my.clevelandclinic.org/health/diseases/22188-chemical-pregnancy}} \textit{"is real and deserves mourning."}
Central to this grief is confronting that one may never have a biological child—whether through failed surgical retrievals\footnote{Surgical sperm retrieval procedure that uses an operating microscope to collect mature sperm for use in in vitro fertilization (IVF) - Source: \url{https://tinyurl.com/y3uranhz}} or the emotional weight of donor gametes.\footnote{Gametes are the reproductive cells (sperm in males and eggs/ova in females) -- Source: \url{https://code-medical-ethics.ama-assn.org/ethics-opinions/gamete-donation}} Yet alongside grief, members affirm that \textit{"our family is of our own making,"} declaring themselves fathers regardless of biological link:
\begin{quote}
``\textit{Deep down, guys often equate masculinity with fertility, so using a donor can feel like it defies our basic programming. That's why partners must validate that we aren't `lesser men' because of a medical issue. I've learned that fatherhood isn't about DNA or seeing my features in my child's face—it's seeing my influence in her character. Guiding her through milestones and modeling how she should be treated is what truly makes her my daughter.}''
\end{quote}

\paragraph{Talking to Family and Friends} 
A recurring pain point is navigating family conversations. Posters describe gatherings as emotional minefields—fielding intrusive questions, insensitive announcements, or comments like \textit{"it's different when you fathered the child."} Those using donor conception describe imposter syndrome as fathers; one admitted \textit{"more anxiety about telling family than actually using donor sperm,"} while another wrote that accepting congratulations \textit{"felt like I was misleading the family"} and baby showers were \textit{"laden with anxiety because I knew it wasn't me who got my wife pregnant."} The pain intensifies with tragedy—one poster described repeatedly explaining to thrilled family, including an ailing parent, that \textit{"there were no heartbeats and the babies were gone,"} turning their support system into \textit{"a source of exhaustion."} Peers strategize about boundary-setting and selective disclosure, advising protection of one's \textit{"sanity first"} even if it means declining events.

\paragraph{Comfort Posting} 
Alongside grief are lighter posts serving as comfort rituals. Members deploy pop culture references to celebrate good results, joke about alternative treatments and past choices now laden with irony, and describe teasing partners \textit{"mercilessly and lovingly"} to weather slumps together. These threads blend humor with encouragement, reframing difficult procedures—from mastering self-administered testosterone injections to facing surgical sperm extractions—as milestones of resilience. Beyond peer support, members actively recommend professional help; one noted that couples therapy \textit{"helped a ton"} and emphasized one need not reach a breaking point to benefit, while another urged a struggling user to see a counselor for better coping strategies.

\paragraph{Connection, Hope, and Celebration} 
The community carves out space for tentative joy while navigating complex emotional boundaries. Ritualized language like \textit{"cautious congrats"} tempers hope with shared recognition of risk, and success stories often appear behind spoiler warnings to protect those not ready to see them. Members celebrate vicariously—one user described another's post-surgery baby as \textit{"the best thing I could've read today,"} finding hope that \textit{"there is still a chance."} and affirmations like \textit{"you guys rock!"} offer reprieve from clinical stressors. Yet users must navigate implicit \textit{"pain olympics"} rules governing whose suffering warrants voice; some newcomers find this emotional intensity intimidating, encountering harsh policing rather than support—a tension we explore by analyzing moderation next. 

\subsubsection{Moderation: Structure, Orientation, and Safety} \label{subsec:mod_qual_findings}
A third theme highlights the importance of moderation and community norms in shaping participation. While these practices are often praised for maintaining safety and coherence, they also produce barriers and frustrations, especially for newcomers.

\paragraph{Onboarding and Private Messaging} 
Onboarding is a blend of humor, empathy, and procedural instruction. Automated greetings often reference the subreddit as \textit{``the best shitty corner of the internet,''} a self-aware attempt to soften the rigidity of rule enforcement. Many new users appreciate this levity, interpreting it as a sign that moderators understand the emotional intensity of infertility. Often newcomers are greeted with an emphatic \textit{"Welcome!"} as well as an invitation to refer to community pinned resources but to not shy away from asking questions. Despite that, some note that such humor does not fully offset the sting of rejection when their posts are removed for not adhering to group rules. In some cases, moderators also encourage members to continue sensitive or off-topic conversations via private messages, a practice that both protects thread coherence and provides space for more personal exchanges. While some valued this redirection as an invitation to connect more deeply, others felt it reinforced exclusion by pushing vulnerable disclosures out of public view. Here again we can see signs of TMI as exemplified by the limited autonomy that new members can express themselves in this subreddit \cite{EnglarCarlson2006,Aguera2017}.


\paragraph{Moderation Policy} 
Often part in parcel of the newcomer experience might be meeting with their posts automodded and removed. Automated bots and moderators enforce strict posting guidelines, regularly reminding users that they enforce keeping to the topic of male infertility; \textit{"reading the room"} is one phrase used to make new posters aware that outside discussions are not welcome. Even basic questions, introductions, or updates are redirected into daily or weekly threads. Automated mods are not the only enforcers of these policies. Seasoned participants remind posters that this subreddit serves men with clinical infertility—not adjacent concerns like low libido, general hormone optimization,\footnote{Maintaining hormones within optimum ranges becomes more difficult with age. Though medically contested \cite{holtorf2009bioidentical}, it is discussed in medical resources: \url{https://nyulangone.org/news/future-you-how-hormones-impact-health-every-age}} or reversing steroid use. Members also distinguish clinical infertility from general \textit{trying to conceive} (TTC): users with normal parameters, or those trying for under a year without severe issues, are redirected to TTC subreddits as their situation lacks the medical and emotional gravity of diagnosed infertility. One moderator drew sharp distinctions: between \textit{"actual issues"} and self-inflicted consequences of steroid use warranting reality checks rather than support; between the \textit{"agony," "guilt and shame"} of men facing infertility versus requests for hormone optimization "coaching"; and between struggling to conceive versus \textit{"odd threads"} about contraception or sexual mechanics. These off-topic discussions, members emphasized, can be \textit{"triggering"} for participants navigating genuine loss—hence their strong discouragement, often with expletives.

Many users express gratitude for this structure, noting that the rules keep the subreddit \textit{``focused and high-signal,''} avoiding the clutter of repetitive or low-effort posts. Others, however, describe the system as alienating when they first join. One poster remarked that their message was \textit{``not visible until a mod approved it,''} leaving them feeling like they were \textit{``posting into a void''} at a moment of emotional urgency. Though the very fact that the rules can be discussed within the forum demonstrate that - to a certain degree - they are flexible and adaptable. In our research, we did encounter users providing feedback on what structures of posts made more sense if \textit{"there wasn't enough traffic for recurring posts"} because they \textit{"are too busy on other parts of Reddit."} Managing the flow of information does seem to be a contested topic open for debate.

\paragraph{Navigation} 
As is evident from the above themes, navigating where the flow of conversations are structure to happen can be a recurrent frustration. Newcomers often post in the wrong place and are redirected with templated bot messages pointing to the FAQ, wiki, or Welcome Wednesday thread. For some, these messages serve as useful orientation tools that provide quick access to resources; others describe them as \textit{``robotic and impersonal,'}' especially when received after disclosing sensitive struggles. A common critique was that automated responses \textit{``felt more like bureaucracy than support,''} creating disorientation during moments of crisis. Yet, once acclimated, many members acknowledge that the structured system of daily treatment chats and thematic threads helps them find relevant discussions and reduces the chaos of a large community. This navigation is not only done using bots, but at times members of the community act as navigators as well. By pointing newcomers to \textit{"the FAQ at the top of the page"} or "pinned posts" that contains the experiential brain trust developed by members over the years, they try to assist in making sure \textit{"most of your questions"} get answered in a timely manner.


\paragraph{r/infertility vs. r/maleinfertility}
Members articulated several perceived differences between r/infertility and \newline r/maleinfertility, though these characterizations were contested. First, demographics: r/infertility was described as female-dominated, where men are \textit{"rare"} and the conversation feels female-focused, while r/maleinfertility offers space for men to discuss how societal emphasis on virility shapes their experience without the \textit{"ban on men showing weakness."} Second, focus: some perceived r/infertility as primarily emotional, with one user finding it \textit{"intimidating"} and another reporting judgmental responses to scientific questions about lifestyle factors. By contrast, r/maleinfertility was characterized as practical and data-driven—members engage in technical discussions of semen analysis parameters and treatment protocols. It is important to note that these distinctions were not universally held. One female member of r/infertility pushed back, clarifying that members there are \textit{"very much interested in the scientific aspects"}—suggesting the emotional-versus-scientific framing oversimplifies both communities.

\subsection{Predictors of More Future Engagement (RQ1b)} \label{sec:future_engagement}
The time-lagged regression analysis presented in Tables 1-3 offers insights into how language use (sentiment), themes discussed by the user, and user engagement behavior on this subreddit community predict higher engagement from month to month in the infertility support community. Specifically, linguistic markers drawn from LIWC (Table \ref{tab:LIWC_future}) show that future engagement and language showing affiliation with others in the community, insight-oriented words, and cognitive processing terms are all positively associated with an increased engagement in the community. Conversely, language associated with detachment or externalization—such as references to money, negation, or fatalistic expressions—is negatively associated with continued participation.

\begin{table}[ht]
\centering
\caption{LIWC coefficients predicting future engagement: Positive (left) vs. negative (right) linguistic marker}
\label{tab:LIWC_future}
\begin{tabular}{lrrrr l lrrrr}
\multicolumn{5}{c}{Positive LIWC} &  & \multicolumn{5}{c}{Negative LIWC} \\ \hline
Feature & Coef & StdErr & t & p &  & Feature & Coef & StdErr & t & p \\ \hline
Affiliation & 0.779 & 0.122 & 6.388 & *** & & Adverb & -0.350 & 0.067   &-5.166 &  ***  \\
 Auxverb&  0.256&  0.075&  3.435&  ***&  &  Article                                  &  -0.377   &  0.059&  -6.348&  ***\\
 Cause&  0.163&  0.039&  4.200&  ***&  &  Death&  -0.049&  0.022&  -2.195&  **\\
 Certain&  0.162&  0.038&  4.237&  ***&  &  Family&  -0.206&  0.055&  -3.743&  ***\\
 Compare&  0.218&  0.043&  5.029&  ***&  &  Female&  -0.110&  0.054&  -2.051&  ***\\
 Conj&  0.207&  0.079&  2.641&  **&  &  Focuspast&  -0.129&  0.048&  -2.682&  ***\\
 Discrep&  0.086&  0.041&  2.092&  **&  &  Home&  -0.048&  0.024&  -2.010&  ***\\
 Ingest&  0.119&  0.046&  2.599&  ***&  &  Male&  -0.261&  0.093&  -2.809&  ***\\
 Insight&  0.320&  0.065&  4.954&  ***&  &  Money&  -0.121&  0.025&  -4.836&  ***\\
 Motion&  0.091&  0.034&  2.673&  ***&  &  Negate&  -0.188&  0.034&  -5.548&  ***\\
 Netspeak&  0.079&  0.034&  2.308&  **&  &  Space&  -0.196&  0.062&  -3.179&  ***\\
 Prep&  0.289&  0.107&  2.708&  ***&  &  Tentative&  -0.155&  0.074&  -2.106&  ***\\
 Risk&  0.121&  0.051&  2.393&  **&  &  We&  -0.381&  0.078&  -4.907&  ***\\
 Sad&  0.214&  0.043&  4.963&  ***&  &  &  &  &  &  \\
 Sexual&  0.072&  0.030&  2.397&  **&  &  &  &  &  &  \\
 Social&  0.301&  0.150&  2.009&  **&  &  &  &  &  &  \\
 Swear&  0.084&  0.029&  2.933&  ***&  &  &  &  &  &  \\
 Time&  0.116&  0.045&  2.578&  ***&  &  &  &  &  &  \\
 Work&  0.131&  0.038&  3.433&  ***&  &  &  &  &  &  \\
\end{tabular}
\end{table}

Complementing the linguistic markers, thematic predictors (Table \ref{tab:topic_label}) show that users who contribute informational and strategic content—such as posts giving \textit{advice on insurance, Doctors, and Medication Expense} and interpreting \textit{Medical Reports}  (see \S\ref{subsec:mecical_advice}) as well \textit{Health and well-being} posts are more likely to remain active in the community over time. In contrast, those whose contributions focus heavily on emotional processing (explained in \S\ref{subsec:findings_emotional_support}) and include \textit{Comfort Posting} or expressions of \textit{Connection, Hope, and Celebration}, \textit{Grief and Disappointment} or \textit{Talking to Family and Friends}), are less likely to re-engage in the future. This may indicate that some users seek temporary emotional relief rather than ongoing participation. However, it might also indicate that social support might be continued on different channels. In our qualitative analysis, we noticed that a number of conversations where users are \textit{``venting''} end in direct messages (DM) to one or more members of the community. They may also move these discussions to virtual meetings  (e.g., Zoom meetings especially in the early days of the pandemic) or in-person meetings. While not focused on emotional processing, the \textit{Female Infertility Generalized Guidance} also decreses the chance of future engagement in the community -- probably because this community is more focused on \textit{\textbf{Male} Infertility Guidance}.

\begin{table}[ht]
\centering
\caption{Thematic predictors of future engagement: Positive (left) vs. negative (right) topic labels}
\label{tab:topic_label}
\begin{tabular}{p{2cm}rrrr l p{2cm}rrrr}
\multicolumn{5}{c}{Positive Theme} &  & \multicolumn{5}{c}{Negative Theme} \\
\hline
Feature & Coef & StdErr & t & p &  & Feature & Coef & StdErr & t & p \\
\hline
 Advice on Insurance, Doctors, and Medication Expense&  0.051&  0.024&  2.136&  ***&  & Comfort Posting&  -0.040&  0.016&  -2.436&  ***\\
 Health and Well-being&  0.061&  0.024&  2.591&  **&  &  Connection, Hope, and Celebration&  -0.065&  0.030&  -2.154&  **\\
 Medical Report Analysis&  0.075&  0.026&  2.873&  ***&  &  Female Infertility Generalized Guidance&  -0.054&  0.024&  -2.283&  **\\
 &  &  &  &  &  &  Grief and Disappointment&  -0.042&  0.018&  -2.380&  **\\
 &  &  &  &  &  &  Talking to Family and Friends&  -0.047&  0.014&  -3.309&  ***\\ \bottomrule 
\end{tabular}
\end{table}

Behavioral metrics from Table \ref{tab:user_meta} reinforce this pattern: longer, more detailed posts and higher post scores (indicating community validation) are strong predictors of future activity, whereas highly positive sentiment or lack of replies correlates with disengagement. However, one surprising find is that the possibility of receiving a response has a negative relationship with the user's continued engagement in the community. This may be related to moving the conversation to other spaces (as discussed above), or to the content of the responses which might demotivate users from future engagement. Finally, negative VADER sentiment is associated with increased future engagement, suggesting that users expressing more negative emotions are more likely to continue participating in the community. This aligns with the emotionally taxing nature of infertility—particularly for men, who often face unique challenges such as stigma, identity threat, limited offline support, and lower interest in male infertility in the scientific community. In this context, negative sentiment is not only expected but may reflect a critical need for support, validation, and shared understanding within a safe, anonymous space.

Taken together, the findings suggest that sustained engagement is driven not simply by emotional need, but by a sense of connection, informational utility, and perceived recognition. Community designers and moderators may foster deeper retention by encouraging bidirectional exchange, prioritizing actionable discussions, and ensuring contributors feel acknowledged and valued.


\begin{table}[H]
\caption{Control variables predicting future engagement: Sentiment, score, comment characteristics, and response probability}
\label{tab:user_meta}
\centering
\begin{tabular}{lrrrr}
\multicolumn{5}{c}{} \\ \hline
Feature & Coef & StdErr & t & p \\ \hline
VADER Sentiment & -0.108 & 0.0276 & -3.904 & *** \\
Score & 0.072 & 0.020 & 3.685 & *** \\
Comment Length & 0.654 & 0.198 & 3.3 & *** \\
Posibility of Response & -0.103 & 0.041 & -2.541 & ** \\
\end{tabular}
\end{table}

\subsection{Community Structure (RQ2a)} \label{sec:cluster_findings} 
After removing clusters with less than 10 users, the Louvain community detection algorithm provided us with eleven micro-communities -- heretofore clusters. As shown in Table 4, all detected clusters exhibit conductance values below 0.5, suggesting that they are structurally cohesive and somewhat insulated from the broader network—a pattern that could indicate potential echo chamber behavior from a purely structural perspective. However, when we apply the second condition for echo chambers proposed by Morini et al. \cite{morini2021toward}—purity greater than 0.7—we find that none of the clusters meet this threshold, with purity values instead remaining low (ranging from 0.07 to 0.16). This indicates that while clusters maintain internal structural coherence, reinforced by moderate to substantial reciprocity, they do not exhibit the high topical or stance homogeneity necessary to be classified as echo chambers. Indeed, the relatively low purity scores, coupled with consistently high topic entropy values (3.9–5.2), suggest that each cluster engages with a wide diversity of themes and stances, a finding elaborated in Appendix \ref{appendix:cluster_analysis} that shows the distribution of stances (defined in \S\ref{subsec:echo_meth}) across different clusters. 

Indeed, we struggled to find thematic signatures for different clusters throughout our qualitative analysis of the clusters. With the exception of smaller clusters like cluster 2 made of 33 users, which focused on  at-home sperm banking, most clusters showed a similar spread of varied topics related to male infertility. Even in that cluster, the discussion diverged to multiple different topics. However, similarities and differences emerged between different roles, and the ways in which they were enacted in different clusters. We discuss this in detail in \S\ref{sec:roles}.

\begin{figure}[ht]
    \centering
    \includegraphics[width=0.9\linewidth]{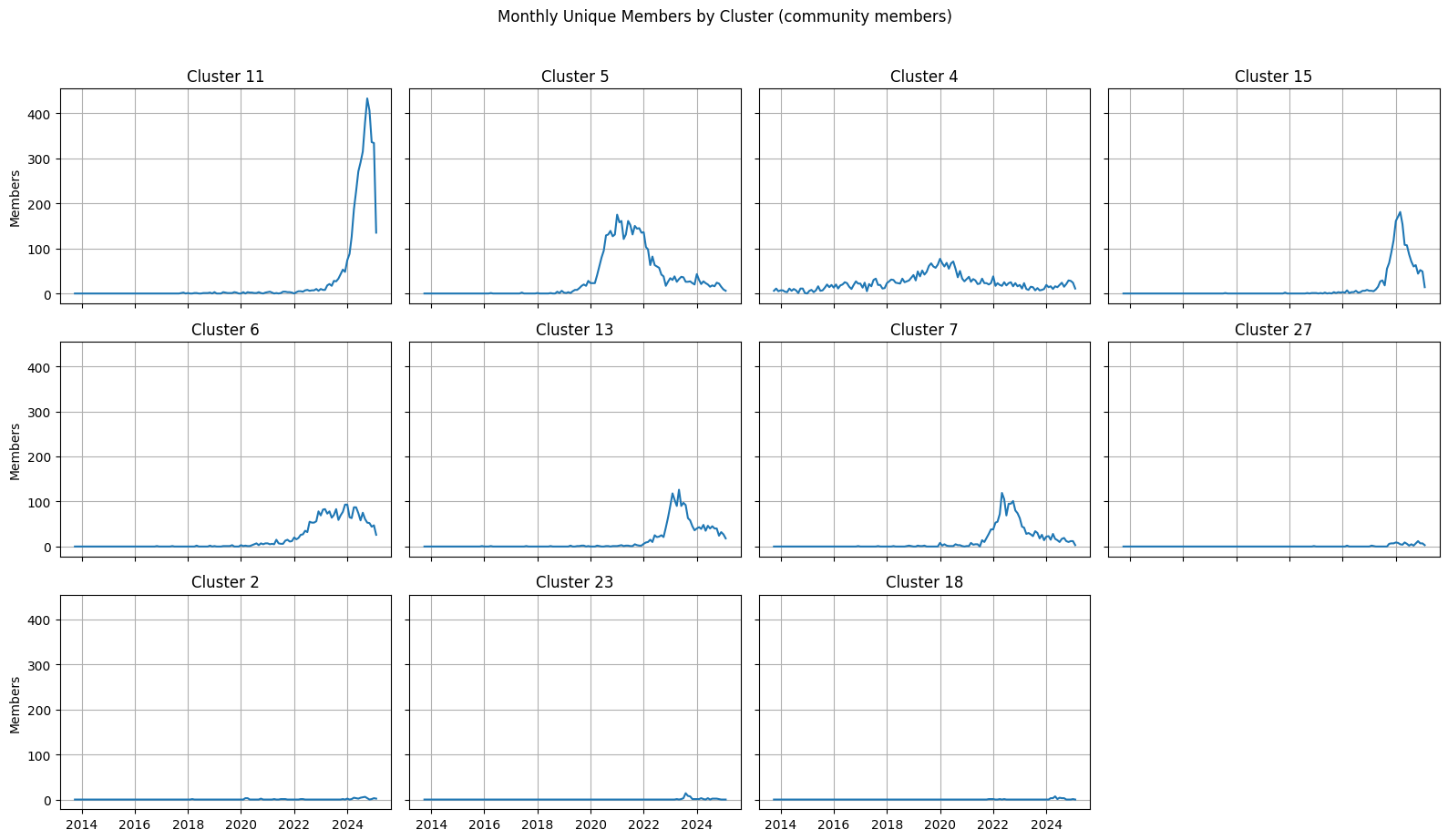}
    \caption{Cluster membership growth over time. Most clusters show steady participation prior to 2020, followed by sharp increases during the COVID-19 lockdown period. This pattern suggests that broader social disruptions acted as catalysts for intensified engagement in male infertility discussions, contributing to the structural consolidation of clusters.}
    \label{fig:cluster_members}
\end{figure}

Finally, as shown in Figure \ref{fig:cluster_members}, most clusters experienced notable growth during the onset of the COVID-19 lockdowns in 2020, suggesting that broader social disruptions served as catalysts for increased participation in male infertility discussions.

\begin{table}[ht]
\centering
\caption{Network cluster characteristics: Structural cohesion, reciprocity, and topical diversity metrics}
\label{tab:community_analysis}
\begin{tabular}{|c|c|c|c|c|l|}
\hline
\textbf{Cluster} & \textbf{Size} & \textbf{Conductance}& \textbf{Reciprocity} &  \textbf{Purity} &\textbf{Topic} 
\\
 &  &  &  &    &\textbf{Entropy} 
\\
\hline
11 & 2,614 & 0.313 & 0.279 &  0.13 &4.907 
\\
\hline
5 & 1,852 & 0.270 & 0.328 &  0.08 &4.829 
\\
\hline
4 & 1,594 & 0.312 & 0.312 &  0.07 &5.232 
\\
\hline
15 & 1,013 & 0.365 & 0.334 &  0.09 &4.667 \\
\hline
6 & 953 & 0.378 & 0.347 &  0.07 &4.926 
\\
\hline
13 & 877 & 0.397 & 0.345 &  0.07 &4.775 
\\
\hline
7 & 800 & 0.378 & 0.346 &  0.09 &4.718 
\\
\hline
27 & 59 & 0.326 & 0.375 &  0.14 &4.522 \\
\hline
2 & 33 & 0.364 & 0.255 &  0.10 &4.568 
\\
\hline
23 & 26 & 0.420 & 0.390 &  0.12 &3.918 
\\
\hline
18 & 15 & 0.222 & 0.346 &  0.16 &3.899 \\
\hline
\end{tabular}
\end{table}

\subsection{Cross-Cluster Roles and Their Variation (RQ2b)} \label{sec:roles}
Certain roles recur across clusters but manifest differently by context. These roles overlap—one poster often wears many hats within a thread, balancing medical information with markers of commiseration. Table \ref{tab:role_emerging} shows role distribution across clusters.
\textit{Mentors}, representing the majority of dominant voices, shepherd new and established members through emotional support and lived-experience advice. In Cluster 7, one mentor detailed expected sperm count recovery thresholds after hormonal therapy; in Cluster 6, another narrated their journey toward acceptance after multiple failed surgeries. Both modeled how to face fertility struggles—the former through experiential medical information, the latter through closure after failure. Mentors acting as hope brokers are nearly universal, sustaining morale by framing setbacks as relatable shared experiences or providing guidance to overcome obstacles. In Clusters 18 and 23, partners emphasized improvements following supplement and vitamin regimens, sustaining optimism even when outcomes remained uncertain.
\textit{Experts} offered detailed instructions on hormone regimens, supplements, and surgical timing. In Cluster 5, a user explained how Clomid influences hormonal pathways; in Cluster 11, another outlined testosterone therapy recovery strategies; in Cluster 15, individuals logged precise dosages and outcomes. Experts provide structured, technical guidance using clinical language, occasionally sprinkled with anecdotal observations.
Both Expert and Mentor roles frequently overlap with \textit{system critical} behavior, as firsthand experience of structural bias, financial burden, and professional blind spots lends weight to their insights. Users argued that fertility clinics marginalize men, criticized doctors pressuring patients into surgery, and warned about conflicts of interest. These communities exist largely because members have faced hardships accessing and affording care while overcoming societal biases regarding male infertility—an undercurrent of systems criticism is unsurprising.

\begin{table}[h!]
\centering
\caption{Distribution of user roles across network clusters}
\label{tab:role_emerging}
\begin{tabular}{|c|p{10cm}|}
\hline
\textbf{Cluster} & \textbf{Roles} \\
\hline
11 & Mentor; Experts; Moderators; Systems Critic; Hope Broker \\
\hline
5 & Mentor; Experts; Systems Critic; Hope Broker \\
\hline
4 & Mentor; Hope Broker; Experts; Moderators \\
\hline
15 & Mentor; Experts; Systems Critic; Hope Broker \\
\hline
6 & Mentor; Experts; Systems Critic; Hope Broker; Newcomer \\
\hline
13 & Mentor; Experts; Systems Critic; Hope Broker \\
\hline
7 & Mentor; Experts; Systems Critic; Hope Broker; Newcomer \\
\hline
27 & Mentor; Experts; Hope Broker; Newcomer \\
\hline
2 & Mentor; Experts; Systems Critic; Hope Broker \\
\hline
23 & Mentor; Experts; Systems Critic; Hope Broker; Newcomer \\
\hline
18 & Mentor; Experts; Systems Critic; Hope Broker \\
\hline
\end{tabular}

\end{table}

\subsubsection{Moderation by Users Across Clusters} \label{subsec:mod_role}
Human moderators and everyday users enforce group norms via explicit forum guidelines. Some apply rules strictly, telling others their results do not qualify for posting. Others soften enforcement, explaining why certain parameters (such as morphology alone) are not meaningful. Still others frame moderation through community response, noting how upvotes reflect shared perceptions of value. Together, these practices show moderation enacted through both automation and social interpretation.

In Cluster 4, we identified norm-setting through empathy, as users cautioned against flaunting good results that might trigger others. Multiple examples showed moderation as scientific correction, with members pushing back against poor advice and urging others to seek specialists. However, Cluster 23—one of the smallest with 26 users—had little formal moderation; extreme claims like abstinence-based fertility hacks went unchecked, indicating smaller clusters may have different moderation styles.
Moderation thus emerges as community guidance balancing rule enforcement with \textit{navigation labor}—often invisible work ensuring newcomers find their footing and discussions remain accessible, high-signal, and emotionally safe. Mentors sometimes overlap with moderators, herding discussions to preserve both coherence and empathy.

\begin{figure}[H]
    \centering
    \includegraphics[width=0.99\linewidth]{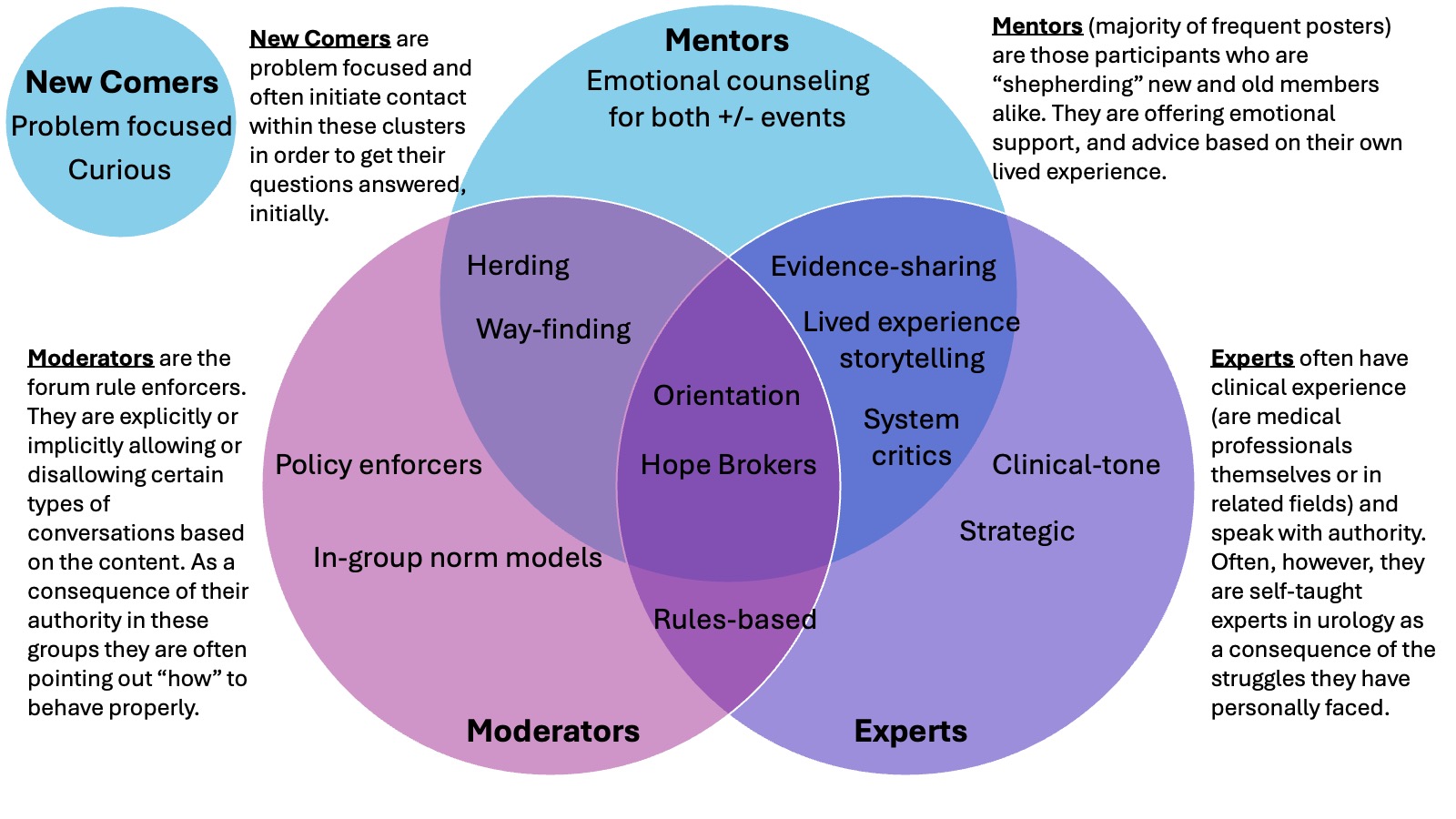}
    \caption{Roles and interactions within male infertility support clusters. The diagram illustrates how newcomers, mentors, moderators, and experts overlap in functions such as way-finding, emotional counseling, policy enforcement, evidence sharing, and strategic medical guidance. These roles collectively shape community navigation, knowledge exchange, and emotional resilience.}
    \label{fig:RolesPic}
\end{figure}

Figure \ref{fig:RolesPic} summarizes identified roles. Mentors and moderators not only enforce posting standards but perform way-finding: directing users to appropriate threads, providing structured resources, and intervening with empathetic language to soften redirects. This shepherding blends procedural guidance with emotional counseling, drawing on lived experience to normalize confusion and ensuring vulnerable disclosures are not left unanswered.

\subsubsection{The Role of Women and Significant Others Across Clusters}
Women and partners are especially visible in clusters where emotional processing and medical distrust dominate. In Cluster 4, women acted as voices of empathy and shared grief—one partner expressed sorrow over a miscarriage and directed others toward supportive communities, affirming shared pain while providing peer guidance.

In Cluster 18, wives frequently narrated their husbands' test results and frustrations with medical encounters. One described her partner's severe semen analysis results; another, as \textit{system critic}, recounted a doctor dismissing a possible connection between varicoceles\footnote{A varicocele is an enlargement of the veins in the scrotum that drain the testicles, which can lead to pooling of blood and may negatively affect male fertility by increasing testicular temperature and impairing sperm production -- Source: \url{https://www.mayoclinic.org/diseases-conditions/varicocele/symptoms-causes/syc-20378771}} and infertility.
In other clusters, partners became \textit{"partner-scientists,"} compiling supplement regimens and interpreting semen analyses on their husbands' behalf. One described how her husband's sperm count rose with lifestyle changes; another detailed an exact supplement stack reportedly improving results. These examples illustrate women's roles as record-keepers, advocates, and interpreters of their partners' medical journeys—turning intimate partnerships into sites of experimentation balancing hope with vigilance.

\subsection{Cross-Posters as Navigators and Mentors in the Male Infertility Community (RQ3)} \label{sec:cross_posting}
\begin{figure}[H]
    \centering
    \includegraphics[width=0.7\linewidth]{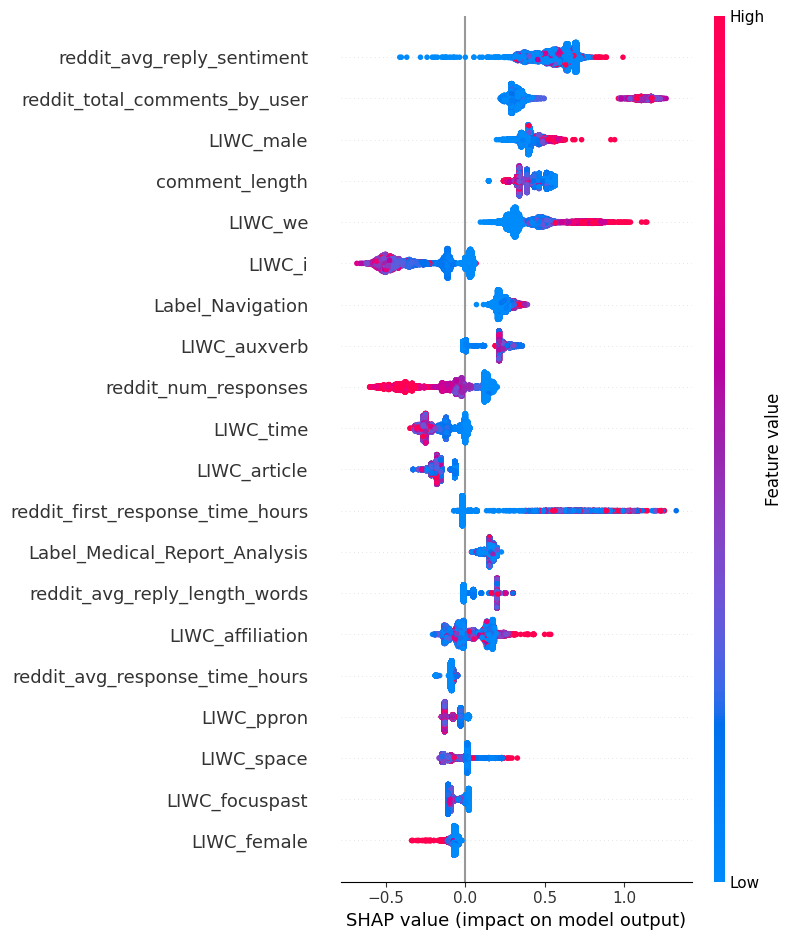}
    \caption{Feature importance for predicting cross-posting behavior. The plot shows both the relative contribution of each feature to the model’s output (SHAP value magnitude) and the direction of influence. Higher values of average reply sentiment, total user comments, and linguistic markers such as male and “we” pronouns increase the likelihood of cross-posting, while self-focused language (“I”), temporal references, and high response volume reduce it.}
    \label{fig:beeswarm}
\end{figure}

The results of classifier predicting cross-posting users who post to both r/maleinfertility and the main gender-mixed r/infertility are visualized using SHAP are presented in Figure \ref{fig:beeswarm}. The details of these predictors are contextualized and explained below.

Cross-posters on Reddit can be characterized by both their activity patterns and linguistic style. They are generally highly active users, contributing more, longer comments than non-cross-posting users, with shorter average response times, and longer, more positive responses, indicating acceptance by other members of the community.

Comments by cross-posters usually focus on \textit{navigation}, where mentors and moderators (see \S\ref{subsec:mod_role}) help users to navigate (see \S\ref{subsec:mod_qual_findings}) the norms of r/maleinfertility. Some of the moderators rely on their knowledge of the normative differences between r/maleinfertility and other subreddits to guide users through its use. In addition to guiding people in community navigation, cross-posting users also guided others in better understanding their medical reports (see \S\ref{subsec:mecical_advice}).

Linguistically, cross-posters shift away from a self-focused voice (“I”) toward a collective or external orientation, using more “we” pronouns and male references while being less involved in female references. They also rely more on narrative or explanatory constructions with auxiliary verbs. These linguistic shifts align with their typical content focus: offering guidance on \textit{navigating} the community and helping others interpret \textit{medical reports}, as discussed above.

Taken together, cross-posters emerge as prolific, engagement-driven users who actively contribute to community support. They not only seek validation by posting across spaces but also provide mentorship and direction to others. Their emphasis on navigation suggests an effort to shape community norms in ways that better accommodate men—potentially filling gaps left unaddressed in other infertility forums.

\section{Discussion}
Our findings portray r/maleinfertility as a hybrid sociotechnical space that simultaneously functions as an informal diagnostic hub, a therapeutic commons, and a governed institution. This configuration helps men navigate the compounded uncertainties of male infertility—biological, relational, and psychological—while also working around Traditional Masculinity Ideologies (TMI) that discourage help-seeking and emotional expression \cite{Messerschmidt2019,Cleary2012,Oliffe2023,Tsan2011}. In contrast to the isolation and “perverse passivity” reported in clinical contexts \cite{dolan_its_2017}, the subreddit affords anonymity and asynchronicity that legitimize disclosure, echoing prior HCI/CSCW work showing that online communities can scaffold identity work and care practices for stigmatized groups \cite{Ammari2024,Randazzo2025,andalibi_2020}. Below, we deepen this account in relation to our research questions (\S\ref{subsec:des_1}) and then translate the results into design (\S\ref{sec:design_rec}) and policy recommendations (\S\ref{sec:policy_imp}).

\subsection{Contributions to Current Literature} \label{subsec:des_1}
We organize our contributions around three analytic dimensions. First, reflecting on RQ1, \S\ref{sec:dis_first} explores how affiliation, insight, and actionable guidance drive sustained engagement through informal diagnostics, logistical coordination, and therapeutic scaffolding. Second, by reflecting on the findings associated with RQ2, \S\ref{sec:disc_second} explore cohesive micro-communities with adaptive, overlapping roles and show how mentorship and moderation extend peer-support and platform-governance accounts. Third, reflecting on RQ3, \S\ref{sec:disc_three} explores how cross-posters normalize norms and transfer know-how across communities, complicating models of knowledge brokering and collective resilience. Together these findings advance theories of social exchange, trauma-informed design, and platform governance by foregrounding the hybrid infrastructures and relational dynamics that sustain participation in stigmatized health contexts.

\subsubsection{Affiliation, insight, and actionable guidance keep people coming back} \label{sec:dis_first}
Based on thematic analysis (115 topics; §\ref{sec:themes}) and time-lagged regression (§\ref{sec:future_engagement}), three discourse arenas emerged—Medical Advice, Emotional Support, and Moderation/Orientation —forming hybrid infrastructure for diagnostics, therapy, and governance. Engagement increased with actionable guidance but declined when users received more replies, countering findings that community responses foster future participation \cite{arguillo_et_al_2006,lampe_et_al_2005,Backstron_et_al_2008,chungWhenPersonalTracking2017,lampe_et_al_2010,Huang_Foote_21}.

\paragraph{Informal diagnostics and logistics: Challenging Traditional Masculinity Ideologies} The dominance of Medical Advice shows how peers translate lab values, treatment options, and system navigation into lay expertise supplementing opaque clinical interactions. This extends work on socially sustained self-tracking and collective sensemaking, where personal data becomes shared, emotionally manageable narratives \cite{10.1145/1753326.1753409,knittel_anyone_2021,mamykina_collective_2015,choe_understanding_2017,luModelSociallySustained2021a,chungWhenPersonalTracking2017,guiWhenFitnessMeets2017}. Social Exchange Theory explains why disclosures persist despite stigma: intrinsic rewards and informational gains outweigh perceived risks \cite{SocialExchangeTheory,wu_effects_2023,yanKnowledgeSharingOnline2016,SocialCapitalIndividual2011}.

Yet technical framing obscures the emotional substrate. Viewing digital practices through a sociomaterial lens—where identities are enacted and negotiated across spaces through material and discursive means \cite{van2011feminist,barad2007meeting}—reveals how members simultaneously challenge and comply with Traditional Masculinity Ideologies (TMI). Much as DIY fathers appropriate traditionally feminine caregiving by reframing it through masculine logics of problem-solving and technical competence \cite{ammari_et_al_2017}, men in r/maleinfertility embed emotional support within medical discourse: decoding semen analyses, comparing protocols, and crowdsourcing interpretations. This allows vulnerability to circulate under the guise of information exchange—legitimizing grief often overlooked offline without explicitly violating TMI norms demanding stoicism \cite{Messerschmidt2019,Cleary2012}.

This explains why members characterized r/infertility as \textit{"female-centered"} and \textit{"emotionally focused"} while viewing r/maleinfertility as data-driven—despite both providing substantial emotional support. The difference lies not in care work's presence but its discursive packaging: r/maleinfertility members performed support through technical registers preserving masculine self-presentation. Dolan et al. \cite{dolan_its_2017} observed similar patterns clinically, where men defined themselves as the "voice of reason" and characterized emotional expression as something women did—limiting disclosure while positioning stoicism as masculine rationality. The medical framing served an analogous function, enabling men to "work around" the "perverse passivity" Dolan identified—where infertility was stoically endured rather than actively addressed—by channeling emotional processing through diagnostic sensemaking.

While this technical framing mirrors women's infertility communities in emphasizing information support \cite{zou2024self} (\S\ref{sec:related_work_carework_ohc}), r/maleinfertility demonstrated robust esteem support, particularly around identity-threatening experiences like donor conception. The insistence that men using donors are not \textit{"lesser men because of a medical issue"} (\S\ref{subsec:findings_emotional_support}) directly addresses the existential crisis Dolan \cite{dolan_its_2017} documented, where infertility "destabilised the ontological connections between a prized masculine identity, the male body and biological fatherhood." Where women's communities offered informational solidarity while avoiding deep relational investment, the male community provided explicit identity repair—reconstructing fatherhood around influence and presence rather than biology.

This resonates with Semaan et al.'s findings on military veterans, where hyper-masculinity learned during service created barriers to post-service disclosure \cite{semaan_et_al_16,semaan_et_al_17}. Veterans who successfully navigated transition often did so by witnessing others disclose online, establishing "a new form of camaraderie not based solely on a hyper-masculine identity" \cite{semaan_et_al_17}. Similarly, r/maleinfertility members observing peers share vulnerable experiences within technical frameworks appeared more willing to disclose—suggesting the medical register provided a TMI-compliant pathway to emotional expression. Hegemonic masculinity operates through cultural ascendancy rather than overt domination \cite{Messerschmidt2019}; the technical discourse may represent a "complicit masculinity" realizing some benefits of masculine norms while enabling care work and esteem support that sustains the community and facilitates identity repair.
\paragraph{Therapeutic scaffolding.} Emotional Support spans grief processing and "comfort posting," aligning with research on digital belonging and men's mental health, where social connection mitigates depression and suicidality \cite{McLaren2009,Bailey2018,HoltLunstad2022,Steptoe2013}. 

Additionally, asynchronous communication fosters well-being \cite{Hull2020,Rahman2022} and supports resilience in trauma-adjacent settings \cite{Ammari2024,Randazzo2025}. Yet our regression complicates the assumption that more replies drive retention: response probability negatively predicts subsequent activity. Possible explanations include overly corrective feedback and off-platform migration, revising participation models in stigmatized contexts \cite{andalibi_2020,ammari_et_al_19}.
While prior work highlights direct peer responses as central to support \cite{Huang_Foote_21}—a tenet of trauma-informed design \cite{chen_trauma-informed_2022, Scott2023}—users may sustain engagement through lurking \cite{randazzo_ammari_2023} or distributing participation across communities \cite{Huang_Foote_21, Ammari2024}. We outline implications in the future work section.

\subsubsection{Cohesive micro-communities with adaptive and overlapping roles} \label{sec:disc_second}
Community detection revealed structurally cohesive clusters below topical purity thresholds—not echo chambers \cite{morini2021toward}—yet the network exhibited notable structural closure (\S\ref{sec:cluster_findings}) with substantial reciprocity, while maintaining topical breadth. This complicates claims that pseudonymous support fosters weak, transactional ties \cite{zou2024self}, and shows that they can provide network support (see \S\ref{sec:related_work_carework_ohc}). However, such closure may paradoxically impede healing given how tight relational networks can produce grief bubbles that trap users in trauma-focused interactions, hindering identity integration even when content remains diverse \cite{Randazzo2025}. The scarcity of male-centric infertility spaces \cite{patel_et_al_2019} may intensify this by concentrating participation into relationally dense communities (we discuss a temporal element to this issue in future work). We next discuss the roles we discovered in this community. 

\paragraph{Moderation as Both Safeguard and Flashpoint}
Wu et al. \cite{wu_et_al_24} show how moderators establish symbolic boundaries differentiating in-group from out-group members. In r/maleinfertility, this manifested in discussions defining who qualifies as clinically infertile versus facing general conception challenges, with the subreddit's \textit{About} section directing partners to dedicated threads. As detailed in \S\ref{subsec:mod_qual_findings}, moderators specified that men with legitimate infertility claims (as opposed to low libido) qualify for membership, though some clusters proved more permissive than others (\ref{sec:cluster_findings}). Unlike fathers-only groups that explicitly excluded mothers \cite{ammari_stayhome_dads_16}, r/maleinfertility's exclusion criteria remained ambiguous. This nuanced boundary-setting at the intersection of gender identities \cite{Gilbert_23} and health conditions \cite{saha_et_al_20} introduced friction that complicated safe space management. For instance, in Cluster 4, users cautioned against flaunting good results that might trigger others—paralleling "pain olympics" among parents of children with special needs \cite{ammari_et_al_2015_networked}. This reflects the contextual nature of online harm \cite{gillespie2018custodians}: even well-intentioned "good news" posts may be discouraged when negative sentiment sustains participation (§\ref{sec:future_engagement}).

\paragraph{Way-Finding and Navigation Labor}
Peer support is central to trauma-informed design \cite{chen_trauma-informed_2022, Scott2023}, and our findings show support extends beyond upvotes and replies \cite{laviolette_using_2019, randazzo_ammari_2023}. Mentors and moderators play critical way-finding roles: contextualizing rules, directing newcomers to resources, and reframing emotional disclosures into actionable steps. This information management—redirecting posts to wikis, daily threads, or pinned resources—fulfills a key moderation function (organizing information \cite{edwards_moderator_2002}) but can frustrate newcomers unfamiliar with community architecture (\S\ref{subsec:mod_qual_findings}). Gilbert \cite{Gilbert_23} demonstrates how such moderation constitutes intensive emotional labor at personal, community, and systemic levels. Building on this, we identify an under-theorized \textit{"navigation labor"} where actors blend rule enforcement with empathetic guidance, shepherding vulnerable disclosures so they are not left unanswered.

\paragraph{Intersecting Roles and Evolving Peer Support}
Roles overlapping with moderation—particularly mentorship—complicate conventional peer support models and reveal moderation as simultaneously protection and collective empowerment \cite{ammari_et_al_22}. Two roles warrant attention: \textit{system critics} and \textit{hope brokers}. System critics—users critiquing fertility clinics for marginalizing men and warning about conflicts of interest—perform what Ammari et al. \cite{ammari_et_al_22} term "empowerment as consciousness" (power within): questioning institutional structures upon finding others facing similar constraints. This criticism reflects material reality: research consistently excludes men from fertility studies despite fertility being couple-based \cite{harlow2020qualitative}, leaving users \textit{"grasping at straws"} for male-factor information (\S\ref{subsec:mecical_advice}).

Hope brokers complement this by enacting "empowerment through community" (power through) \newline
—members gaining resources and efficacy through experienced peers \cite{ammari_et_al_22}. By framing setbacks as relatable and highlighting treatment gains, they sustain optimism when outcomes remain uncertain, mirroring "networked empowerment" among parents of children with special needs \cite{ammari_et_al_2015_networked}. Together, system critics raising consciousness and hope brokers modeling pathways create conditions for "empowerment in community" (power with): collective action addressing common concerns \cite{ammari_et_al_22}. The overlapping roles (Figure \ref{fig:RolesPic}) reveal users wearing "many hats" within single threads—distributed governance building collective capacity to challenge inadequate care systems and hegemonic masculinity \cite{Messerschmidt2019,dolan_its_2017}. Understanding peer support evolution is central to trauma-informed design (\S\ref{sec:related_work_trauma_informed}).

\subsubsection{Bridge-builders who normalize norms and transfer know-how} \label{sec:disc_three}
When answering RQ3, we found cross-posters resemble collective-oriented brokers: they write generalizable, guidance-seeking content, lean on “we” pronouns, and often activate when engagement is limited locally—complementing prior observations of cross-community knowledge brokering and empathy building especially when managing trauma \cite{Ammari2024,Randazzo2025}. This finding also complicates the study of norm creation at different community levels (i.e., The Reddit platform $\rightarrow$  several related subreddits $\rightarrow$  one subreddit $\rightarrow$  micro-communities within the subrreddit) \cite{Chandrasekharan_et_al_2018} as it shows how cross-posters can engage in norm-setting in different communities relying on their deep knowledge of community norms in both spaces. Additionally, as we showed in \S\ref{subsec:mod_qual_findings}, the norms differences between r/maleinfertility and r/infertility are contested, which raises questions about how these differences are understood over time and how cross-posters affect the evolution of both communities.

\subsection{Design Implications} \label{sec:design_rec}
Grounded in the above, we reframe design not as content-topic recommendation but as \emph{role-aware, journey-sensitive infrastructure} that supports safe visibility, effective way-finding, and portable care.

\subsubsection{Role-transportability: Recommending Users Based on their Assumed Roles}
Earlier work highlighted narrative transportability—providing users access to relevant content—offering design approaches to help people discover new communities and connect around shared topical interests \cite{randazzo_ammari_2023,Randazzo2025,Ammari2024}.

We extend this idea to role transportability. Building on role embeddings that model structural similarity in networks (e.g., identifying users who bridge subgraphs) \cite{ahmed2020role,ahmed2019role2vec}, we incorporate temporal dynamics \cite{grayson2019temporal} and behavioral signals from discourse and interaction patterns \cite{saxena2022users,akar2025connecting}. This hybrid approach produces role-based embeddings enabling recommendation systems to match newcomers with experienced members who have played relevant roles over time. Consistent with user agency (a central tenet of trauma-informed design as presented in \S\ref{sec:user_choice_voice}) and affirmative consent design principles, individuals must retain the option to opt out of such recommendations \cite{im_et_al_2021}. This recommendation system would move from suggesting people who talked about X theme to “people who \underline{act} like \textit{mentors/experts/seekers/hope brokers}. Rather than pushing users into a single cluster by topic, the system would \emph{situationally} connect them to interactional neighborhoods that match their current role needs (e.g., a \textit{new comer} is surfaced to active \textit{mentors} who recently provided high-quality diagnostic scaffolding). This better reflects how users likely “become part of a cluster” through \emph{interaction} rather than content alone.

\subsubsection{Therapist-ready, user-controlled summaries of on-platform engagement}
Our data suggest that some disclosures move to DMs or off-platform, with many replies encouraging therapy (\S\ref{subsec:findings_emotional_support}). To support continuity of care and reduce the emotional burden of “retelling” \cite{bagwell2015intimate}, we propose an opt-in system that summarizes and anonymizes a user’s recent online engagements (e.g., threads joined, key concerns, decisions), with editable and redactable outputs. These summaries could be exported to Fast Healthcare Interoperability Resources or other therapist–client media, complementing mixed therapy modalities such as two-way text communication \cite{Hull2020}.

\subsubsection{Visible role and expertise tags—with optional professional verification}
Building on our role findings in \S\ref{sec:roles} and prior work on badges, awards, and verified participation \cite{burtch2022peer,trujillo2022assessing,bornfeld2017gamifying} in \S\ref{sec:user_choice_voice}, we recommend a multilayered tagging and verification system to enhance visibility and trust within the community. 

Community-earned role tags—such as Mentor, Navigator, Systems-wise, or Report-Interpreter—would be driven by behavioral signals and community nominations. These tags should be revocable and scoped to specific contexts so that recognition reflects sustained, relevant engagement rather than one-time actions.

Complementing these role tags, members could voluntarily adopt topic or skill tags (e.g., insurance navigation, azoospermia recovery) to make their expertise legible to seekers. In parallel, professionally verified accounts for clinicians and counselors would strengthen credibility, following verification approaches such as license checks and scoped disclaimers proposed in \cite{Ammari2024}. Importantly, these accounts must include clear guardrails to prevent their presence from being misconstrued as a substitute for clinical care.



\subsection{Policy Implications.} \label{sec:policy_imp}
Male infertility communities on Reddit reveal critical gaps in formal care: fathers often receive
inadequate medical guidance and lack access to psychological support, fueling the system critic role that challenges medical and policy shortcomings. By formally recognizing moderation as care work, platforms and public health agencies should resource moderator training for health stigma contexts (trauma-informed scripts, de-escalation, misinformation triage). This acknowledges the dual governance/therapeutic role highlighted in our findings. Secondly, policy makers and health advocates alike could go further to bring awareness to male infertility in public health discourse. Campaigns and clinical guidelines should explicitly address men’s barriers under TMI, normalizing help-seeking and couple-oriented care pathways \cite{Messerschmidt2019,dolan_its_2017,Seidler2016}. Additionally, by mandating that mental health care be considered part of fertility care much of the stigma associated with that care as well as access to those invaluable resources could be made more available to this community of patients. As we’ve demonstrated, embed screening and referral to online/offline support alongside treatment planning, leveraging evidence that connection and belonging mitigate risk in men’s mental health \cite{McLaren2009,Bailey2018,HoltLunstad2022}.

\section{Limitations and Future Work}
This study faces several limitations. First, while time-lagged regression revealed that receiving replies can sometimes reduce the likelihood of future engagement, we are not yet certain \textit{why} this occurs. One possibility is that conversations migrate to private messages or off-platform spaces, but it is also plausible that the tone or content of responses discourages sustained participation. This ambiguity calls for future qualitative research—particularly interviews with community members and moderators—to better understand the dynamics of responses and their impact on engagement.

Second, moderation emerged as both contentious and necessary. Automated enforcement and human moderators play essential roles in ensuring high-quality discourse, yet they are also described as alienating, especially by newcomers. The dual nature of moderation—as governance and as care work—warrants deeper exploration. Future studies could directly engage with moderators across infertility communities to map their challenges, strategies, and needs.

Finally, our network analysis provides only a snapshot of cluster structures. Modeling how clusters evolve over time, including their topical diversity and survival rates of users within them, would allow for a richer understanding of community trajectories. Longitudinal modeling could reveal how echo chamber risks, mentorship roles, and cross-posting practices shift during different stages of the infertility journey. Such work would deepen our grasp of how digital infertility communities adapt to changing membership, external disruptions (e.g., COVID-19 lockdowns), and evolving cultural discourses around masculinity and reproduction.

\section{Conclusion}
Male infertility communities on Reddit operate as informal diagnostic hubs, therapeutic commons, and governed spaces that help men navigate stigma and uncertainty. Role differentiation—especially mentors and moderators as way-finders—enables collective sensemaking while balancing care and rule enforcement. Linguistic and topical signals tied to affiliation and actionable guidance predict continued engagement, whereas some replies may inadvertently deter public return. Clusters are cohesive yet not echo chambers. Designing for navigation labor, empathetic moderation, and cross-community bridges can make support more inclusive and sustainable.

\bibliographystyle{ACM-Reference-Format}
\bibliography{sample-base}

\newpage

\section{Appendices}

\appendix

\section{Research Methods}
\subsection{Topic Modeling}
\label{appendix:topic_model}
\subsubsection{Appendix: Hyperparameter Selection}
\label{appendix:hyperparameter}

The hyperparameter search focused on three key parameters:

\begin{enumerate}
    \item \textbf{UMAP Neighborhood Size}: The \textbf{Uniform Manifold Approximation and Projection (UMAP)} algorithm \cite{mcinnes2018umap} is used to reduce the dimensionality of the document embeddings before clustering. We varied the \textit{n\_neighbors} parameter across \{20, 30, 40, 50\}, which determines the number of neighboring sample points considered when constructing the manifold approximation.
    \begin{itemize}
        \item \textbf{Smaller values} (e.g., 20) result in a more \textbf{local view}, capturing fine-grained structure and enabling smaller, tightly packed clusters.
        \item \textbf{Larger values} (e.g., 50) promote a \textbf{more global view}, favoring larger, more cohesive clusters.
        \item Adjusting this parameter helps balance local topic specificity and global topic cohesion.
    \end{itemize}

    \item \textbf{Minimum Topic Size}: The \textit{min\_topic\_size} parameter controls the smallest allowable topic size in the clustering process \cite{grootendorst_bertopic_2022}. We explored values of \{20, 30, 40, 50\}:
    \begin{itemize}
        \item \textbf{Lower values} (e.g., 20) allow for the formation of \textbf{many smaller topics}, potentially increasing topic granularity.
        \item \textbf{Higher values} (e.g., 50) enforce the creation of \textbf{larger, more general topics}, ensuring stability but possibly reducing topic diversity.
        \item Setting this parameter appropriately is crucial, particularly for large datasets where higher values (e.g., 100–500) may be necessary to avoid excessive topic fragmentation.
    \end{itemize}

    \item \textbf{N-gram Range}: The \textit{n\_gram\_range} parameter was varied between (1,2) and (1,3), affecting how words are tokenized in the topic representation using \textbf{CountVectorizer} \cite{pedregosa2011scikit}.
    \begin{itemize}
        \item \textbf{(1,2) n-grams} allow for single words and \textbf{bi-grams}, capturing common phrase structures such as ``climate change.''
        \item \textbf{(1,3) n-grams} further include \textbf{tri-grams}, enabling more nuanced topic representations such as ``machine learning model.''
        \item This adjustment helps capture multi-word expressions and improves the semantic coherence of extracted topics.
    \end{itemize}
\end{enumerate}

\subsubsection{Appendix: Model Evaluation}

Each of the \textbf{32 model configurations} (combining different \textit{UMAP neighborhood sizes, minimum topic sizes, and n-gram ranges}) was trained and evaluated using the \textbf{u\_mass coherence score}, a measure of topic quality that relies on word co-occurrence probabilities \cite{newman2010automatic}. Coherence scoring is widely used in topic modeling to assess how semantically meaningful the generated topics are. The configuration yielding the highest coherence score was selected as the \textbf{optimal set of parameters} for final topic modeling.

\begin{figure}[H]
    \centering
    \includegraphics[width=0.9\linewidth]{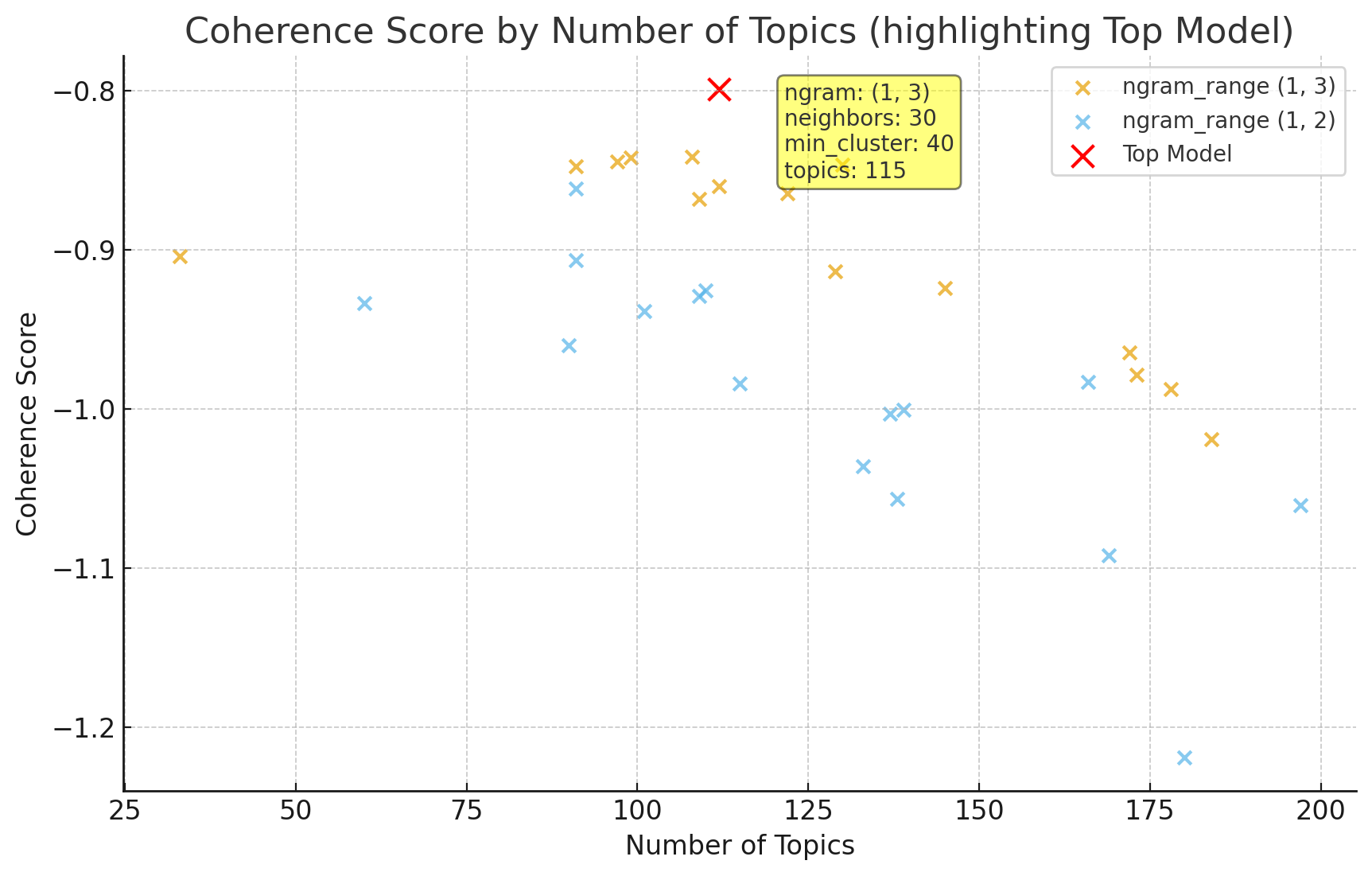}
    \caption{Figure shows the spread of topic models along with associated topics and coherence scores. The top model with the highest coherence score is highlighted along with its parameters.}
    \label{fig:topiccoherence}
\end{figure}

\subsection{Topic Breakdown}
\label{appedix:topic_breakdown}
\subsubsection{Appendix: Medical Advice Analysis}

\begin{figure}[H]
    \centering
    \includegraphics[width=1.0\linewidth]{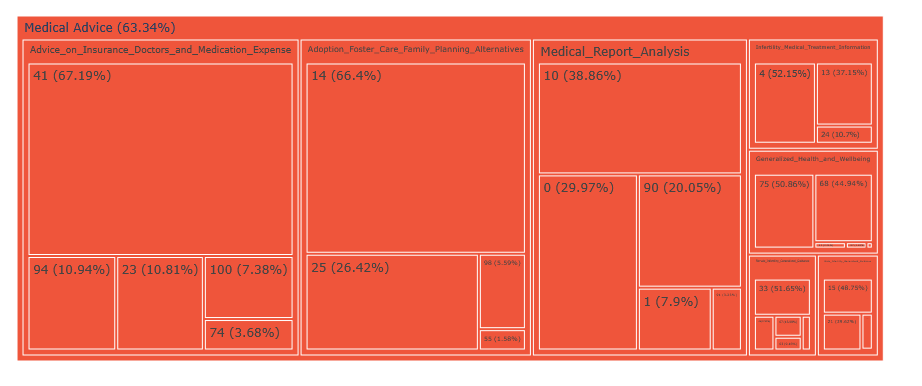}
    \caption{Hierarchical treemap of Medical Advice discussions (63.34\% of topics). Subthemes include advice on insurance, doctors, and medication expenses; adoption and family planning alternatives; interpretation of medical reports; fertility treatment protocols; and generalized health and wellbeing. This cluster shows how community members act as lay interpreters of biomedical knowledge while navigating systemic and financial barriers.}
    \label{fig:treemap_medical_advice}
\end{figure}

The Medical Advice cluster (63.34\% of all discussion topics) represents the largest and most active domain of infertility discourse. It encompasses detailed exchanges on diagnostics, treatment planning, systemic barriers, and adjunct strategies. The treemap (Figure~\ref{fig:treemap_medical_advice}) shows how topics are distributed across major sub-clusters:

\paragraph{1. Advice on Insurance, Doctors, and Medication Expense} 
\begin{itemize}
    \item \textbf{Key topics: 41, 94, 23, 100, 74}
    \item Discussions foreground financial and logistical barriers to care. Users share tips for contesting insurance denials, finding affordable clinics or pharmacies, and coping with high out-of-pocket costs. 
    \item Topic 41 focuses on accessing and managing fertility medications, where patients compare discounts, shared-risk programs, and international pharmacies. 
    \item These discussions reveal how medical advice is inseparable from the economic realities of fertility care.
\end{itemize}

\paragraph{2. Adoption, Foster Care, and Family Planning Alternatives} 
\begin{itemize}
    \item \textbf{Key topics: 14, 25, 98, 35}
    \item Users weigh non-biological pathways such as adoption, donor eggs, and foster care. 
    \item Topic 14 emphasizes the emotional, ethical, and identity dilemmas of using donor gametes, while Topic 25 covers the decision-making process between IUI and IVF. 
    \item These posts articulate grief over lost genetic continuity alongside hope for alternative family-building.
\end{itemize}

\paragraph{3. Medical Report Analysis} 
\begin{itemize}
    \item \textbf{Key topics: 10, 0, 90, 1}
    \item This cluster centers on interpreting semen analyses, hormone panels, and embryo grading. 
    \item Topic 10 covers deep dives into semen analysis and male factor infertility, while Topic 0 focuses on navigating varicocele and testicular treatments. 
    \item Topic 90 addresses treatment comparisons (IUI vs. IVF). Collectively, these illustrate how Reddit users act as peer translators of medical data, filling interpretive gaps left by clinicians.
\end{itemize}

\paragraph{4. Fertility, Medical Treatment Information} 
\begin{itemize}
    \item \textbf{Key topics: 4, 13, 24}
    \item Focused on protocol-level knowledge, this sub-cluster includes embryo transfer preparation, ovulation induction, and questions about endometrial thickness. 
    \item Posters blend clinical metrics with emotional uncertainty, often sharing experiences of failed cycles and marginal results.
\end{itemize}

\paragraph{5. Generalized Health and Wellbeing} 
\begin{itemize}
    \item \textbf{Key topics: 75, 68, 33, 18, 21}
    \item Highlights lifestyle interventions and everyday health practices. 
    \item Discussions include supplements (CoQ10, L-carnitine, antioxidants), exercise, diet, and abstinence from alcohol/porn. 
    \item While some advice is biomedical, much is experiential and anecdotal, reflecting a culture of self-experimentation alongside medical treatment.
\end{itemize}

\paragraph{Summary} 
This treemap illustrates the dominance of Medical Advice in infertility discourse, emphasizing how community members function as lay interpreters, navigators, and strategists. Medical data (semen analyses, hormone results) are contextualized through peer translation, while treatments are weighed against cost, accessibility, and systemic barriers. The presence of adoption and lifestyle sub-clusters shows how members also grapple with broader existential and self-care dimensions.

\subsubsection{Appendix: Emotional Support Treemap Analysis}
\begin{figure}
    \centering
    \includegraphics[width=1.0\linewidth]{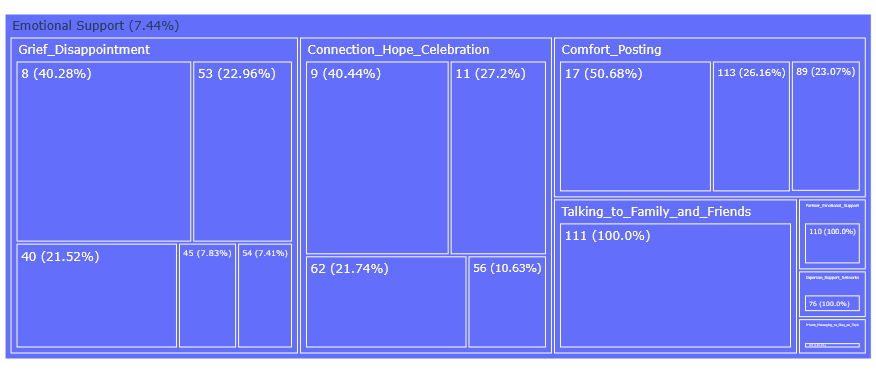}
    \caption{Hierarchical treemap of Emotional Support discussions (7.44\% of topics). Subthemes include grief and disappointment, connection and celebration, comfort posting, talking to family and friends, and externalized support networks (e.g., pets, online ties). This cluster highlights the community’s role as a therapeutic space for collective grieving, resilience, and coping.}
    \label{fig:treemap_emotional_support}
\end{figure}

The Emotional Support cluster (7.44\% of all discussion topics) illustrates how the infertility community functions as a therapeutic scaffold, offering empathy, coping strategies, and collective resilience. The treemap (Figure~\ref{fig:treemap_emotional_support}) shows how topics are organized into several major sub-clusters:

\paragraph{1. Grief and Disappointment} 
\begin{itemize}
    \item \textbf{Key topics: 8, 53, 40, 45, 54}
    \item Threads capture raw expressions of grief following miscarriages, failed IVF or IUI cycles, and prolonged infertility. 
    \item Members describe feelings of being ``numb'' or ``broken,'' often framing infertility as ambiguous loss. 
    \item Community responses include validation and refrains of solidarity (e.g., ``you are not alone''), creating a shared lexicon of care.
\end{itemize}

\paragraph{2. Connection, Hope, and Celebration} 
\begin{itemize}
    \item \textbf{Key topics: 9, 11, 62, 56}
    \item Discussions focus on moments of optimism: positive pregnancy tests, supportive clinic staff, or milestones achieved during treatment. 
    \item Humor, cautious celebration, and expressions of hope help balance the heavier narratives of grief. 
    \item These threads highlight how small wins become rituals of resilience.
\end{itemize}

\paragraph{3. Comfort Posting} 
\begin{itemize}
    \item \textbf{Key topics: 17, 113, 89}
    \item Emotional check-ins where users seek comfort and reassurance. 
    \item Posts often serve as informal therapy, allowing members to vent frustrations, share daily struggles, and receive encouragement. 
    \item Community validation reinforces a sense of belonging during vulnerable moments.
\end{itemize}

\paragraph{4. Talking to Family and Friends} 
\begin{itemize}
    \item \textbf{Key topic: 111}
    \item Focused on relational dynamics, this cluster captures the difficulty of sharing infertility experiences with family, friends, or partners. 
    \item Strategies include boundary-setting around pregnancy announcements, selective disclosure, or managing insensitive remarks. 
    \item These discussions underscore how infertility reshapes social relationships and communication practices.
\end{itemize}

\paragraph{5. Externalized Support Networks} 
\begin{itemize}
    \item \textbf{Key topics: 110, 76, 6}
    \item Beyond family and peers, members discuss support from pets, online networks, and community rituals. 
    \item Pets are frequently described as emotional anchors, providing comfort and companionship when treatment outcomes disappoint. 
    \item These forms of externalized support highlight creative coping in the face of medical and social uncertainty.
\end{itemize}

\paragraph{Summary} 
The Emotional Support treemap illustrates the psychosocial dimension of infertility communities. Members engage in collective grieving, celebrate small victories, and co-create rituals of resilience. Through comfort posting and discussions of relational strain, the community offers not only emotional validation but also practical coping strategies for navigating stigma and isolation.

\subsubsection{Appendix: Moderation Treemap Analysis}

\begin{figure}
    \centering
    \includegraphics[width=1.0\linewidth]{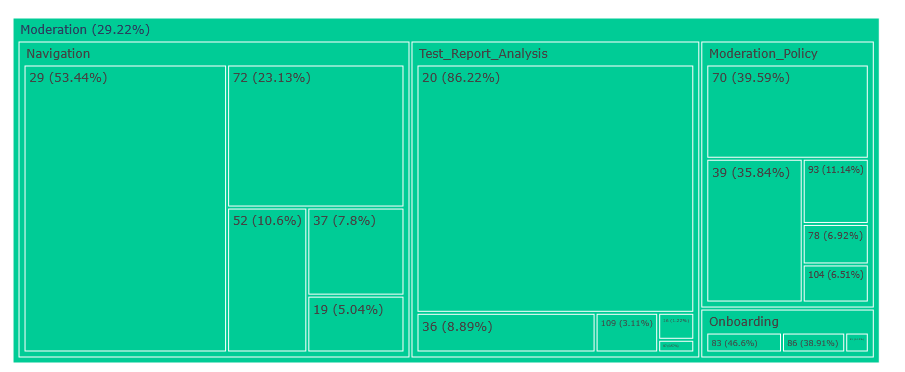}
    \caption{Hierarchical treemap of Moderation discussions (29.22\% of topics). Subthemes include navigation and rule guidance, test report posting standards, policy enforcement, and onboarding rituals. This cluster illustrates the community’s governance infrastructure, showing how moderators and bots maintain order, safeguard emotional safety, and structure knowledge exchange.}
    \label{fig:treemap_moderation}
\end{figure}

The Moderation cluster (29.22\% of all discussion topics) highlights the digital governance and infrastructural practices that sustain community order, safety, and knowledge accessibility. The treemap (Figure~\ref{fig:treemap_moderation}) shows how moderation-related discussions are organized into several sub-clusters:

\paragraph{1. Navigation and Rule Guidance} 
\begin{itemize}
    \item \textbf{Key topics: 29, 72, 52, 37, 19}
    \item These discussions focus on helping users find their way through the community’s rules, FAQs, and posting guidelines. 
    \item Members are often redirected to wiki pages, daily threads, or pinned posts when their questions or updates overlap with existing resources. 
    \item Navigation ensures high signal-to-noise ratio and maintains a structured flow of information, though it can sometimes frustrate newcomers.
\end{itemize}

\paragraph{2. Test Report Analysis and Posting Standards} 
\begin{itemize}
    \item \textbf{Key topics: 20, 36}
    \item Moderators enforce rules around posting diagnostic data, such as semen analysis results or medical reports. 
    \item Bots and community guides provide templates for interpreting results, helping members avoid panic over borderline values. 
    \item These practices reflect trauma-informed moderation, balancing empathy with accuracy and order.
\end{itemize}

\paragraph{3. Moderation Policy and Enforcement} 
\begin{itemize}
    \item \textbf{Key topics: 70, 39, 93, 78, 104}
    \item A cluster dedicated to enforcing participation guidelines and managing inappropriate posts. 
    \item Discussions cover redirecting updates to designated threads, discouraging repetitive questions, and maintaining tone. 
    \item This ensures respectful, high-quality discourse while preserving emotional safety for participants.
\end{itemize}

\paragraph{4. Onboarding and Community Orientation} 
\begin{itemize}
    \item \textbf{Key topics: 83, 86, 3}
    \item New members are welcomed through automated but empathetic messages that introduce both rules and community culture. 
    \item Orientation blends practical guidance with humor, signaling that the space is simultaneously regulated and supportive. 
    \item This fosters early belonging while reinforcing boundaries and expectations.
\end{itemize}

\paragraph{Summary} 
The Moderation treemap illustrates the invisible labor of governance that underpins the infertility community. Through navigation guidance, posting standards, policy enforcement, and onboarding rituals, moderators and automated systems co-create a structured environment. This infrastructure not only protects informational quality but also enhances psychosocial safety, enabling the community to function as a reliable health and support resource.

\newpage

\section{Cluster Community Analysis}
\label{appendix:cluster_analysis}
Appendix B presents the detailed stance-coding scheme referenced in \S\ref{subsec:echo_meth}. Tables 6-8 show each cluster defined in \S\ref{sec:cluster_findings} along with the stances defined as themes (defined in \S\ref{sec:themes}) crossed with VADER sentiment. 

\begin{table}[htbp]
\centering
\caption{Stance distribution across clusters: Top 10 weighted theme-sentiment combinations by cluster}
\label{tab:community_topic_analysis}
\begin{tabular}{|c|p{12cm}|}
\hline
\textbf{Cluster} & \textbf{Top 10 Stances (with their weights)}\\
\hline
11 &  1. Navigation (+ve) -- 15.2\% \newline
2. Adoption Foster Care Family Planning Alternatives (+ve) -- 12.1\% \newline
3. Medical Report Analysis (+ve) -- 10.6\% \newline
4. Advice on Insurance Doctors and Medication Expense (+ve) -- 9.1\% \newline
5. Advice on Insurance Doctors and Medication Expense (-ve) -- 5.3\% \newline
6. Adoption Foster Care Family Planning Alternatives (-ve) -- 5.3\% \newline
7. Navigation (-ve) -- 5.3\% \newline
8. Generalized Health and Wellbeing (+ve) -- 3.8\% \newline
9. Grief Disappointment (neu) -- 3.0\% \newline
10. Grief Disappointment (-ve) -- 3.0\%     \\
\hline
5 & 1. Advice on Insurance Doctors and Medication Expense (+ve) -- 13.7\% \newline
2. Medical Report Analysis (+ve) -- 10.0\% \newline
3. Adoption Foster Care Family Planning Alternatives (+ve) -- 9.7\% \newline
4. Navigation (+ve) -- 9.4\% \newline
5. Advice on Insurance Doctors and Medication Expense (neu) -- 7.5\% \newline
6. Advice on Insurance Doctors and Medication Expense (-ve) -- 4.8\% \newline
7. Test Report Analysis (+ve) -- 4.8\% \newline
8. Navigation (neu) -- 4.4\% \newline
9. Generalized Health and Wellbeing (neu) -- 4.1\% \newline
10. Adoption Foster Care Family Planning Alternatives (-ve) -- 4.0\%  \\
\hline
18 &  
1. Medical Report Analysis (+ve) -- 20.0\% \newline
2. Advice on Insurance Doctors and Medication Expense (+ve) -- 16.4\% \newline
3. Grief Disappointment (neu) -- 7.3\% \newline
4. Adoption Foster Care Family Planning Alternatives (neu) -- 5.5\% \newline
5. Navigation (neu) -- 5.5\% \newline
6. Advice on Insurance Doctors and Medication Expense (neu) -- 3.6\% \newline
7. Grief Disappointment (+ve) -- 3.6\% \newline
8. Medical Report Analysis (-ve) -- 3.6\% \newline
9. Comfort Posting (-ve) -- 3.6\% \newline
10. Health and Wellbeing (+ve) -- 3.6\% \\
\hline

\hline
\end{tabular}
\end{table}

\begin{table}[htbp]
\centering
\caption{Stance distribution across clusters: Top 10 weighted theme-sentiment combinations by cluster (continued)}
\label{tab:community_topic_analysis_2}
\begin{tabular}{|c|p{12cm}|}
\hline
\textbf{Cluster} & \textbf{Top 10 Stances (with cluster weights)}\\

\hline
4 &  
1. Advice on Insurance Doctors and Medication Expense (+ve) -- 12.9\% \newline
2. Adoption Foster Care Family Planning Alternatives (+ve) -- 11.0\% \newline
3. Medical Report Analysis (+ve) -- 10.8\% \newline
4. Navigation (+ve) -- 9.6\% \newline
5. Advice on Insurance Doctors and Medication Expense (-ve) -- 5.1\% \newline
6. Test Report Analysis (+ve) -- 4.7\% \newline
7. Adoption Foster Care Family Planning Alternatives (-ve) -- 4.5\% \newline
8. Moderation Policy (+ve) -- 3.9\% \newline
9. Navigation (-ve) -- 3.1\% \newline
10. Advice on Insurance Doctors and Medication Expense (neu) -- 3.0\% \\

\hline
15 & 1. Advice on Insurance Doctors and Medication Expense (+ve) -- 14.5\% \newline
2. Medical Report Analysis (+ve) -- 10.5\% \newline
3. Adoption Foster Care Family Planning Alternatives (+ve) -- 9.9\% \newline
4. Navigation (+ve) -- 9.2\% \newline
5. Advice on Insurance Doctors and Medication Expense (neu) -- 6.1\% \newline
6. Test Report Analysis (+ve) -- 5.8\% \newline
7. Advice on Insurance Doctors and Medication Expense (-ve) -- 5.6\% \newline
8. Adoption Foster Care Family Planning Alternatives (-ve) -- 3.6\% \newline
9. Adoption Foster Care Family Planning Alternatives (neu) -- 3.4\% \newline
10. Navigation (-ve) -- 3.1\% \\
\hline
6 & 1. Adoption Foster Care Family Planning Alternatives (+ve) -- 13.8\% \newline
2. Advice on Insurance Doctors and Medication Expense (+ve) -- 11.0\% \newline
3. Medical Report Analysis (+ve) -- 8.8\% \newline
4. Navigation (+ve) -- 6.5\% \newline
5. Advice on Insurance Doctors and Medication Expense (-ve) -- 5.9\% \newline
6. Test Report Analysis (+ve) -- 5.7\% \newline
7. Adoption Foster Care Family Planning Alternatives (-ve) -- 5.3\% \newline
8. Advice on Insurance Doctors and Medication Expense (neu) -- 4.5\% \newline
9. Adoption Foster Care Family Planning Alternatives (neu) -- 3.8\% \newline
10. Moderation Policy (+ve) -- 2.9\%  \\
\hline
13 & 
1. Advice on Insurance Doctors and Medication Expense (+ve) -- 11.2\% \newline
2. Medical Report Analysis (+ve) -- 10.5\% \newline
3. Adoption Foster Care Family Planning Alternatives (+ve) -- 9.8\% \newline
4. Test Report Analysis (+ve) -- 8.2\% \newline
5. Navigation (+ve) -- 7.9\% \newline
6. Advice on Insurance Doctors and Medication Expense (neu) -- 4.9\% \newline
7. Advice on Insurance Doctors and Medication Expense (-ve) -- 4.5\% \newline
8. Adoption Foster Care Family Planning Alternatives (-ve) -- 4.4\% \newline
9. Navigation (-ve) -- 4.0\% \newline
10. Adoption Foster Care Family Planning Alternatives (neu) -- 3.4\% \\

\hline
\end{tabular}
\end{table}

\begin{table}[htbp]
\centering
\caption{Stance distribution across clusters: Top 10 weighted theme-sentiment combinations by cluster (continued)}
\label{tab:community_topic_analysis_2}
\begin{tabular}{|c|p{12cm}|}
\hline
\textbf{Cluster} & \textbf{Top 10 Stances (with cluster weights)}\\

\hline
7 &  
1. Advice on Insurance Doctors and Medication Expense (+ve) -- 14.9\% \newline
2. Medical Report Analysis (+ve) -- 9.6\% \newline
3. Navigation (+ve) -- 9.2\% \newline
4. Adoption Foster Care Family Planning Alternatives (+ve) -- 9.1\% \newline
5. Test Report Analysis (+ve) -- 6.5\% \newline
6. Advice on Insurance Doctors and Medication Expense (-ve) -- 5.8\% \newline
7. Advice on Insurance Doctors and Medication Expense (neu) -- 5.3\% \newline
8. Adoption Foster Care Family Planning Alternatives (neu) -- 4.1\% \newline
9. Adoption Foster Care Family Planning Alternatives (-ve) -- 4.1\% \newline
10. Navigation (-ve) -- 3.8\% \\
\hline
27 & 1. Advice on Insurance Doctors and Medication Expense (+ve) -- 20.7\% \newline
2. Adoption Foster Care Family Planning Alternatives (+ve) -- 12.6\% \newline
3. Medical Report Analysis (+ve) -- 7.0\% \newline
4. Navigation (+ve) -- 6.3\% \newline
5. Advice on Insurance Doctors and Medication Expense (-ve) -- 6.0\% \newline
6. Moderation Policy (+ve) -- 5.6\% \newline
7. Test Report Analysis (+ve) -- 4.6\% \newline
8. Advice on Insurance Doctors and Medication Expense (neu) -- 4.6\% \newline
9. Adoption Foster Care Family Planning Alternatives (neu) -- 3.2\% \newline
10. Adoption Foster Care Family Planning Alternatives (-ve) -- 3.2\% \\
\hline
2 & 1. Navigation (+ve) -- 15.2\% \newline
2. Adoption Foster Care Family Planning Alternatives (+ve) -- 12.1\% \newline
3. Medical Report Analysis (+ve) -- 10.6\% \newline
4. Advice on Insurance Doctors and Medication Expense (+ve) -- 9.1\% \newline
5. Advice on Insurance Doctors and Medication Expense (-ve) -- 5.3\% \newline
6. Adoption Foster Care Family Planning Alternatives (-ve) -- 5.3\% \newline
7. Navigation (-ve) -- 5.3\% \newline
8. Health and Wellbeing (+ve) -- 3.8\% \newline
9. Grief Disappointment (neu) -- 3.0\% \newline
10. Grief Disappointment (-ve) -- 3.0\% \\
\hline
23 & 1. Adoption Foster Care Family Planning Alternatives (+ve) -- 12.3\% \newline
2. Test Report Analysis (+ve) -- 12.3\% \newline
3. Medical Report Analysis (+ve) -- 8.9\% \newline
4. Navigation (+ve) -- 8.2\% \newline
5. Navigation (neu) -- 8.2\% \newline
6. Advice on Insurance Doctors and Medication Expense (+ve) -- 8.2\% \newline
7. Advice on Insurance Doctors and Medication Expense (neu) -- 6.8\% \newline
8. Advice on Insurance Doctors and Medication Expense (-ve) -- 4.8\% \newline
9. Test Report Analysis (-ve) -- 3.4\% \newline
10. Navigation (-ve) -- 3.4\% \\ \hline

\end{tabular}
\end{table}

\end{document}